\documentclass[twocolumn,showpacs,aps,epsfig]{revtex4}

\usepackage{graphicx}
\usepackage{epstopdf}
\usepackage{latexsym}

\usepackage[center]{subfigure}

\begin{document}

 \newcommand{\bq}{\begin{equation}}
 \newcommand{\eq}{\end{equation}}
 \newcommand{\bqn}{\begin{eqnarray}}
 \newcommand{\eqn}{\end{eqnarray}}
 \newcommand{\nb}{\nonumber}
 \newcommand{\lb}{\label}
\newcommand{\PRL}{Phys. Rev. Lett.}
\newcommand{\PL}{Phys. Lett.}
\newcommand{\PR}{Phys. Rev.}
\newcommand{\CQG}{Class. Quantum Grav.}

\title{Spacetime singularities in string and its low dimensional effective theory}
\author{Preet Sharma $^{1}$}
%
\author{Andreas Tziolas $^{1}$}
%
\author{Anzhong Wang$^{1, 2}$}
\email{anzhong_wang@baylor.edu}
\author{Zhong Chao Wu $^{2}$}
\email{zcwu@zjut.edu.cn}
\affiliation{  
$^{1}$ GCAP-CASPER, Physics Department, Baylor University,
Waco, TX 76798-7316\\
$^{2}$ Department of Physics, Zhejiang University of Technology, Hangzhou, 
310032, China}

\date{\today}

\begin{abstract}

Spacetime singularities are studied in both  the $D+d$-dimensional  string 
theory and its $D$-dimensional effective theory, obtained by the Kaluza-Klein   
compactification. It is found that spacetime singularities in the low dimensional 
effective theory may or may not remain after lifted to the $D+d$-dimensional  
string theory, depending on particular solutions. It is also found that there exist
cases in which spacetime singularities appearing in high/low dimensional 
spacetimes do not necessarily happen on the same surfaces.
  
\end{abstract}

\vspace{.7cm}


\maketitle

\vspace{1.cm}

\section{Introduction}

\renewcommand{\theequation}{1.\arabic{equation}}
\setcounter{equation}{0}

String and M-theory all suggest that we may live in a world that 
has more than three spatial dimensions \cite{string1,string2,string3}. Because only three of these are 
presently observable, one has to explain why the others are hidden from 
detection.  One such explanation is the so-called Kaluza-Klein (KK) 
compactification, according to which the
size of the extra dimensions is very small (often taken to be on the order
of the Planck length) \cite{zcw1,zcw2}.  As a consequence, modes that have momentum in the
directions of the extra dimensions are excited at currently inaccessible
energies.
 
In such a frame of compactifications, one often finds that the low-dimensional 
effective theories are usually singular, and such singularities might disappear
when we consider the problem in the original high dimensional spacetimes.
In particular, it was shown that the 4-dimensional extreme black hole with a
special dilaton coupling constant can be interpreted as a completely non-singular,
non-dilatonic, black p-brane in (4+p)-dimensions, provided that $p$ is
odd  \cite{GHT95}.

Recently, three of the current authors \cite{TWW08} studied the problem in the frame 
of colliding two timelike branes in string theory. After developing the general 
formulas to describe such events, we studied a particular class of exact solutions 
first in the 5-dimensional effective theory, and then lifted it to the 10-dimensional 
spacetime. In general, the 5-dimensional spacetime is singular, due to the mutual 
focus of the two 
colliding 3-branes. Non-singular cases also exist, but with the price that both of the
colliding branes violate all the three energy conditions, weak, dominant, and strong
\cite{HE73}.
After lifted to 10 dimensions,  we found that the spacetime remains singular,
whenever it is singular in the 5-dimensional effective theory. In the cases where
no singularities are formed after the collision, we found that the two 8-branes necessarily
violate all the energy conditions. 

In this paper we shall address the same problem, but for the sake of simplicity,
we consider only the case where no branes are present.  Specifically, the
paper is organized as follows. In the next section, we will set up the model to be
studied in the framework of both $D+d$-dimensional  string theory and its $D$-dimensional 
effective theory. In Section III, we  study two classes of exact solutions, by paying
particular attention on their local and global singular behavior. In Section IV,
we first lift these solutions to the $D+d$-dimensional spacetimes of string theory, and 
then study their singular behavior. Section V contains our conclusions and
discussing remarks.

\section{Toroidal Compactification of the Effective Action}

\renewcommand{\theequation}{2.\arabic{equation}}
\setcounter{equation}{0}

Let us consider the toroidal compactification of the NS-NS sector of the action 
 in (D+d) dimensions, $\hat{M}_{D+d} = M_{D}\times M_{d}$, where for the string 
 theory we have
$D+d = 10$. Then, the action takes the form
\cite{LWC001,LWC002,LWC003},
\bqn
\lb{2.1}
S_{D+d} &=& - \frac{1}{2\kappa^{2}_{D+d}}
\int{d^{D+d}x\sqrt{\left|\hat{g}_{D+d}\right|}  e^{-\hat{\Gamma}} \left\{
{\hat{R}}_{D+d}[\hat{g}]\right.}\nb\\
& & \left. + \hat{g}^{AB}\left(\hat{\nabla}_{A}\hat{\Gamma}\right)
\left(\hat{\nabla}_{B}\hat{\Gamma}\right) - \frac{1}{12}{\hat{H}}^{2}\right\},
\eqn
where in this paper we consider the $(D+d)$-dimensional spacetimes described by the
metric,
\bqn
\lb{2.2}
d{\hat{s}}^{2}_{D+d} &=& \hat{g}_{AB} dx^{A}dx^{B} = 
   \gamma_{\mu\nu}\left(x^{\lambda}\right) dx^{\mu}
dx^{\nu} \nb\\& & + \hat{\Phi}^{2}\left(x^{\lambda}\right) 
\hat{\gamma}_{ab}\left(z^{c}\right)dz^{a} dz^{b},
\eqn
with  $\gamma_{\mu\nu}\left(x^{\lambda}\right)$ and $\hat{\Phi}^{2}
\left(x^{\lambda}\right)$ depending only on the coordinates  $x^{\lambda}$ of 
the spacetime $M_{D}$, and $\hat{\gamma}_{ab}\left(z^{c}\right)$ only on the internal 
coordinates $x^{c}$, where $\mu, \nu, \lambda = 0, 1, 2, ...,
D-1$; $a, b, c = D, D+1, ..., D+d - 1$; and $A, B, C = 0, 1, 2, ...,
D+ d -1$. Assuming that matter fields are all independent of $z^{a}$, 
one finds that the internal space $M_{d}$ must be Ricci flat $R[\hat{\gamma}] = 0$ 
\cite{LWC001,LWC002,LWC003}. For the purpose of the current work, it is sufficient to assume 
that $M_{d}$ is a $d-$dimensional torus, $T^{d} = S^{1} \times
S^{1} \times ... \times S^{1}$.  
Then, we find that 
\bqn
\lb{2.3}
\hat{R}_{D+d}\left[\hat{g}\right] &=& R_{D}\left[{\gamma}\right] 
+ \frac{d(d-1)}{\hat{\Phi}^{2}} 
\gamma^{\mu\nu}\left(\nabla_{\mu}\hat{\Phi}\right)\left(\nabla_{\nu}\hat{\Phi}\right)\nb\\
& & - \frac{2}{\hat{\Phi}^{d}}\gamma^{\mu\nu} \left(\nabla_{\mu}\nabla_{\nu}\hat{\Phi}^{d}\right).
\eqn
Ignoring the dilaton  $\hat{\Gamma}$ and the form field  $\hat{H}$, 
the integration of the action (\ref{2.1})
over the internal space yields,
\bqn
\lb{2.4}
S_{D}^{eff.} &=& - \frac{1}{2\kappa^{2}_{D}} \int{d^{D}x\sqrt{\left|\gamma\right|}
 \hat{\Phi}^{d}\left\{R_{D}\left[\gamma\right] \right.}\nb\\
& & \left. 
+ \frac{d(d-1)}{\hat{\Phi}^{2}} 
\gamma^{\mu\nu}\left(\nabla_{\mu}\hat{\Phi}\right)\left(\nabla_{\nu}\hat{\Phi}\right)
\right\},
\eqn
where 
\bq
\lb{2.5}
\kappa_{D}^{2} \equiv \frac{\kappa_{D+d}^{2}}{V_{s}},
\eq
and $V_{s}$ is defined as
\bq
\lb{2.6}
V_{s} \equiv \int{\sqrt{\hat{\gamma}} d^{d}z}.
\eq
For a string scale compactification, we have $V_{s} = 
\left(2\pi \sqrt{\alpha'}\right)^{d}$, where $\left(2\pi  \alpha'\right)$
is the inverse string tension.

After the conformal transformation,
\bq
\lb{2.7}
g_{\mu\nu} = \hat{\Phi}^{\frac{2d}{D-2}}\gamma_{\mu\nu},
\eq
the D-dimensional effective action of Eq.(\ref{2.4}) can be cast in the
minmally coupled form,
\bqn
\lb{2.8}
S_{D}^{eff.} &=&  - \frac{1}{2\kappa^{2}_{D}}\int{d^{D}x\sqrt{\left|g_{D}\right|}
  \left\{ R_{D}\left[g\right] 
-  \kappa^{2}_{D} \left(\nabla\phi\right)^{2}\right\}},\nb\\ 
\eqn
where 
\bq
\lb{2.9}
\phi \equiv \pm 
  \left(\frac{(D+d-2)d}{\kappa^{2}_{D}\left(D-2\right)}\right)^{\frac{1}{2}} 
  \ln\left(\hat{\Phi}\right).
\eq

In this paper, we refer the actions of Eqs. (\ref{2.1}) and (\ref{2.4})  to as  {\em the string frames},
and the one of Eq.(\ref{2.8})   {\em the Einstein frames}. It should be noted that
solutions related by this conformal transformation can have completely different 
physical and goemetrical properties in the two frames. In particular, in one 
frame a solution can be singular, while in the other it can be totally free from 
any kind of singularities. A simple example is the conformally-flat spacetimes
$g_{AB} = \Omega^{2}(x) \eta_{AB}$, where the spacetime described by $g_{AB}$
can have a completely different spacetime structure from that of the Minkowski,
$\eta_{AB}$.

Although we are mainly interested in the string theory with the split
$D = 5$ and $d = 5$, in most cases considered in this paper 
we shall not impose these restrictions here,
so that our results obtained in this paper can be applied to other situations.

\section{Solutions in D-dimensional Spacetimes in   the   Einstein Frame}

\renewcommand{\theequation}{3.\arabic{equation}}
\setcounter{equation}{0}


The variation of the action (\ref{2.8}) with respect to $g_{\mu\nu}$ and
$\phi$ yields the D-dimensional Einstein-scalar field equations,
\bqn
\lb{3.1a}
& & R_{\mu\nu} = \kappa^{2}_{D} \phi_{,\mu} \phi_{,\nu},\\
\lb{3.1b}
& & \nabla_{\lambda} \nabla^{\lambda} \phi = 0,
\eqn
where   $(\;)_{,\mu} \equiv
\partial (\;)/\partial x^{\mu}$.

In this paper, we consider the D-dimensional spacetimes described by the metric
\bq
\lb{3.2}
ds^{2}_{D,E} = 2 e^{2\sigma(u, v)}du dv - e^{2h(u, v)} d\Sigma^{2}_{D-2},
\eq
where 
\bq
\lb{3.3}
d\Sigma^{2}_{D-2} \equiv \sum_{i = 2}^{D-1}{\left(dx^{i}\right)^{2}}.
\eq
Clearly, the $(D-2)$-dimensional space ${\cal{S}}$, spaned by  $d\Sigma^{2}_{D-2}$, 
is Ricci flat, and its topology remains unspecified. In this paper, we assume that 
this space is compact. One example is that it is a $(D-2)$-dimensional torus.

It should be noted that metric (\ref{3.2}) is invariant under the coordinate
transformation,
\bq
\lb{3.8}
u = f(\bar{u}), \;\;\;\; v = g(\bar{v}),
\eq
where $f(\bar{u})$ and $g(\bar{v})$ are arbitrary functions of their indicated
arguments. 

Introducing the following two null vectors \cite{Wang1,Wang2,Wang3,Wang4}, 
\bqn
\lb{3.8a}
l_{\mu} &\equiv& \frac{\partial u}{\partial x^{\mu}} = \delta^{u}_{\mu},\nb\\
n_{\mu} &\equiv& \frac{\partial v}{\partial x^{\mu}} = \delta^{v}_{\mu}, 
\eqn
we can see that both of them are future-directed and orthogonal to ${\cal{S}}$.
In addition, each of these two null vectors defines an affinely parameterized
null geodesic congruence,
\bq
\lb{3.8b}
l^{\lambda}\nabla_{\lambda}l_{\mu} = 0 = n^{\lambda}\nabla_{\lambda}n_{\mu}.
\eq
Then, the expansions of the null ray $u = Const.$ and the one $v = Const.$
are defined, respectively, by
\bqn
\lb{3.8c}
\theta_{l}   &\equiv& \nabla^{\lambda}l_{\lambda} 
                = \left(D-2\right)e^{-2\sigma}h_{,v},\nb\\
\theta_{n}   &\equiv& \nabla^{\lambda}n_{\lambda} 
                = \left(D-2\right)e^{-2\sigma}h_{,u}. 
\eqn
It should be noted that the two null vectors are uniquely defined only upto a
factor \cite{HE73,Wang1,Wang2,Wang3,Wang4}.  In fact,  
\bq
\lb{3.8ca}
\bar{l}_{\mu} = f(u)\delta^{u}_{\mu},\;\;\;
\bar{n}_{\mu} = g(v)\delta^{v}_{\mu},
\eq
represent another set of null vectors that also define affinely parameterized null 
geodesics, and  the corresponding expansions
are given by 
\bq
\lb{3.8cb}
\bar{\theta}_{+} = f(u)\theta_{+},\;\;\;
\bar{\theta}_{-} = g(v)\theta_{-}. 
\eq
However, since along each curve $u = Const.\; \left(v = Const.\right)$, the function 
$f(u) \;\left(g(v)\right)$ is constant, this does not affect the definition of 
trapped surfaces in terms of the expansions (See \cite{HE73,Wang1,Wang2,Wang3,Wang4} in details). Thus, 
without loss of generality, in the following definitions of trapped surfaces and 
black holes we consider only the expressions given by Eq.(\ref{3.8c}).

{\em Definitions} \cite{Pen68,Hay941,Hay942,Wang1,Wang2,Wang3,Wang4}:  The spatial $(D-2)$-dimensional 
surface ${\cal{S}}$ of constant $u$ and $v$
is said {\em trapped, marginally trapped, or untrapped}, according to whether
$\left. \theta_{l}\theta_{n}\right|_{{\cal{S}}} > 0$,
$\; \left. \theta_{l}\theta_{n}\right|_{{\cal{S}}} = 0$,
or $\left. \theta_{l}\theta_{n}\right|_{{\cal{S}}} < 0$.
Assuming that on the marginally trapped surfaces ${\cal{S}}$ we have
$\left.\theta_{l}\right|_{{\cal{S}}} = 0$, then an {\em apparent horizon} is
the closure $\tilde{\Sigma}$ of a three-surface $\Sigma$ foliated by the trapped
surfaces $\cal{S}$ on which $\left.\theta_{n}\right|_{\Sigma} \not= 0$.
It is said {\em outer, degenerate, or inner}, according to
whether $\left.{\cal{L}}_{n}\theta_{l}\right|_{\Sigma} < 0$,
$\left.{\cal{L}}_{n}\theta_{l}\right|_{\Sigma} = 0$, or
$\left.{\cal{L}}_{n}\theta_{l}\right|_{\Sigma} > 0$, where ${\cal{L}}_{n}$
denotes the Lie derivative along the normal direction
${n}_{\mu}$, given by,
\bqn
\lb{3.8d}
{\cal{L}}_{n}\theta_{l}   &=& \left(D-2\right)e^{-4\sigma}
\left(h_{,uv} - 2\sigma_{,u}h_{,v}\right),\nb\\
{\cal{L}}_{l}\theta_{n}   &=& \left(D-2\right)e^{-4\sigma}
\left(h_{,uv} - 2\sigma_{,v}h_{,u}\right). 
\eqn
In addition, if $\left. \theta_{n}\right|_{\Sigma} < 0$
then the apparent horizon is said {\em future}, and if
$\left. \theta_{n}\right|_{\Sigma} > 0$ it is said {\em past}.

{\em Black holes} are usually defined by the existence of {\em future outer
apparent horizons} \cite{HE73,Hay941,Hay942,Ida00,Wang1,Wang2,Wang3,Wang4}. However, in a definition given by
Tipler \cite{Tip77} the degenerate case was also included.

For the metric (\ref{3.2}), we find that Eqs.(\ref{3.1a}) and (\ref{3.1b})  yield
\bqn
\lb{3.4a}
& &  h_{,uu} + {h_{,u}}^{2} - 2 h_{,u}\sigma_{,u} = 
            - \frac{\kappa^{2}_{D}}{D-2}{\phi_{,u}}^{2},\\
\lb{3.4b}
& & h_{,vv} + {h_{,v}}^{2} - 2 h_{,v}\sigma_{,v} = 
           - \frac{\kappa^{2}_{D}}{D-2}{\phi_{,v}}^{2},\\
\lb{3.4c}
& &  2\sigma_{,uv} + (D-2)\left(h_{,uv} + {h_{,u}}h_{,v}\right) = 
  -  \kappa^{2}_{D} \phi_{,u}\phi_{,v},\nb\\
\\
\lb{3.4d}
& &   h_{,uv} + (D-2)h_{,u}h_{,v} = 0,\\
\lb{3.4e}
& &  2\phi_{,uv} + (D-2)\left(h_{,u}\phi_{,v} + h_{,v}\phi_{,u}\right) = 0.
\eqn
It should be noted that Eqs.(\ref{3.4a})-(\ref{3.4e}) are not all  
independent. In fact,
Eq.(\ref{3.4c}) is the integrability condition of Eqs.(\ref{3.4a}) and (\ref{3.4b}), and
can be obtained from Eqs.(\ref{3.4a}), (\ref{3.4b}) and (\ref{3.4e}). Therefore,  
the field equations reduce to Eqs. (\ref{3.4a}), (\ref{3.4b}), (\ref{3.4d}) and 
(\ref{3.4e}). To find the solution, one may first integrate Eq.(\ref{3.4d}) to 
find $h$, which gives,
\bq
\lb{3.5}
h(u, v)  =  \frac{1}{D-2}\ln\left(F(u) + G(v)\right),
\eq
where $F(u)$ and $G(v)$ are arbitrary functions. Then, one can integrate 
Eq.(\ref{3.4e}) to find $\phi$. Once $h$ and $\phi$ are found, $\sigma$ can be 
obtained by integrating  Eqs.(\ref{3.4a}) and (\ref{3.4b}). The general solutions
for $\sigma$ and $\phi$ are unknown. In the following, we shall consider some
specific solutions. In particular, we shall consider the three cases separately: 
(a) $F'(u) \not= 0,\; G'(v)  = 0$; 
(b) $F'(u) = 0,\; G'(v)  \not= 0$; and  (c) $F'(u) G'(v)  \not= 0$, where a prime
denotes the ordinary differentiation.  The second case can be obtained from the first
one by exchanging the $u$ and $v$ coordinates. Thus, without loss of
generality, we need consider only Cases (a) and (c). 

Before proceesing further, we note that for the solution of Eq.(\ref{3.5}), 
 Eqs.(\ref{3.8c}) and (\ref{3.8d}) reduce to
\bqn
\lb{3.5a}
\theta_{l}   &=&  e^{-2\sigma}\frac{G'(v)}{F(u) + G(v)},\nb\\
\theta_{n}   &=& e^{-2\sigma}\frac{F'(u)}{F(u) + G(v)},
\eqn
and 
\bqn
\lb{3.5b}
{\cal{L}}_{n}\theta_{l}   &=& - \theta_{l}\left(\theta_{n} 
+ 2e^{-2\sigma}\sigma_{,u}\right),\nb\\ 
{\cal{L}}_{l}\theta_{n}   &=& - \theta_{n}\left(\theta_{l} 
+ 2e^{-2\sigma}\sigma_{,v}\right). 
\eqn

\subsection{ $F'(u) \not= 0,\; G'(v)  = 0$}

In this case, from Eq.(\ref{3.4b}) we find that $\phi = \phi(u)$. Hence, Eq.(\ref{3.4c}) 
yields
\bq
\lb{3.6}
\sigma(u,v) = a(u) + b(u),
\eq
where $a(u)$ and $b(u)$ are other arbitrary functions. Using the gauge
freedom of Eq.(\ref{3.8}), without loss of generality we can always set 
$a(u) = 0 = b(u)$, so that this class of solutions are given by
\bqn
\lb{3.7}
\sigma(u,v) &=& 0,\nb\\
h(u,v) &=& \ln\alpha(u),\nb\\
\phi(u,v) &=& \pm \sqrt{\frac{D-2}{\kappa^{2}_{D}}}
\int^{u}{\left(- \frac{\alpha''(u')}{\alpha(u')}\right)^{1/2}
du'} \nb\\
& & + \phi_{0},
\eqn
where $\alpha(u) \equiv F(u)^{1/(D-2)}$, and $\phi_{0}$ is an integration constant.
Inserting Eq.(\ref{3.7}) into Eq.(\ref{3.5a}), we find that
\bq
\lb{3.7a}
\theta_{l}   = 0,\;\;\;  
\theta_{n}   = \left(D-2\right)\frac{\alpha'(u)}{\alpha(u)},
\eq
for which we have $\theta_{l} \theta_{n} = 0$ identically. Then, according the above
definition, the $(D-2)$-surfaxe ${\cal{S}}$ is   always marginally trapped. Since
\bqn
\lb{3.7b}
{\cal{L}}_{n}\theta_{l}   =  0 =  {\cal{L}}_{l}\theta_{n}, 
\eqn
${\cal{S}}$ is also degenerate.  

To study these solutions further, we notice that 
for these solutions  all the scalars built from the Riemann 
curvature tensor are zero, therefore, in the present case scalar curvature singularities
are always absent \cite{ES77}.  However, non-scalar curvature singularities might also exist. 
In particular, tidal forces experienced by an observer may become infinitely large under
certain conditions \cite{HWW02}. To see how this can happen, let us consider the timelike 
geodesics in the $(u, v)$-plane, which in the present case are simply given by
\bq
\lb{3.9}
\dot{u} = \gamma_{0},\;\;\;\;
\dot{v} = \frac{1}{2\gamma_{0}},\;\;\;\;
\dot{x}^{i} = 0,
\eq
where $i = 2, ..., D-1$, $\; \gamma_{0}$ is an integration  constant, 
and an overdot denotes the
ordinary derivative with respect to the proper time, $\lambda$, of the timelike
geodesics. Defining $e^{\mu}_{(0)} = d{x}^{\mu}/d\lambda$, we find that the  
unit vectors, given by
\bqn
\lb{3.10}
e^{\mu}_{(0)} &=& \gamma_{0}\delta^{\mu}_{u} +
\frac{1}{2\gamma_{0}}\delta^{\mu}_{v},\nb\\
e^{\mu}_{(1)} &=& \gamma_{0}\delta^{\mu}_{u} -
\frac{1}{2\gamma_{0}} \delta^{\mu}_{v},\nb\\
e^{\mu}_{(i)} &=&   \frac{1}{\alpha(u)} \delta^{\mu}_{i}, 
\eqn
form a freely falling frame, 
\bq
\lb{3.11}
e^{\mu}_{(\alpha)}e^{\nu}_{(\beta)} g_{\mu\nu} = \eta_{\alpha\beta},\;\;\;
e^{\mu}_{(\alpha); \nu}e^{\nu}_{(0)}  = 0,
\eq
where $\eta_{\alpha\beta} = {\mbox{diag.}}\;\{- 1, \;  1, ..., \;  1\}$.
Projecting the Ricci tensor onto the above frame, we find that
\bqn
\lb{3.12}
R_{(\alpha)(\beta)} &\equiv& R_{\mu\nu}e^{\mu}_{(\alpha)}e^{\nu}_{(\beta)} \nb\\
&=& - \gamma^{2}_{0}(D-2)  \left(\frac{\alpha''(u)}{\alpha(u)}\right) 
\left(\delta^{u}_{\alpha}\delta^{u}_{\beta} \right.\nb\\
& & \left. -
\left(\delta^{u}_{\alpha}\delta^{v}_{\beta} 
+ \delta^{v}_{\alpha}\delta^{u}_{\beta}\right)
+ \delta^{v}_{\alpha}\delta^{v}_{\beta}\right).
\eqn
Clearly,  the tidal forces remain finite over the whole spacetime, as long as 
$\alpha''/\alpha$ is finite. To see this clearly, let us consider the following
solutions,
\bq
\lb{3.13} 
 \frac{\alpha''(u)}{\alpha(u)}  = 
- \frac{\omega^{2}}{(u - u_{0})^{\gamma}},
\eq
for which we have
\bqn
\lb{3.15}
\phi(u,v) &=& \phi_{0}
   + \left(\frac{\omega^{2}(D-2)}{\kappa^{2}_{D}}\right)^{1/2}
\nb\\
& & \times \cases{\frac{2}{2-\gamma}(u - u_{0})^{1 - \gamma/2}, & $ \gamma \not= 2$,\cr
\ln{(u - u_{0})}, &  $ \gamma = 2$,\cr}
\eqn
 where $u_{0}$ is an arbitrary constant, and without
loss of generality, we can always set $u_{0} = 0$, an assumption we shall adopt in the 
following discussions. The constants $\omega$ and $\gamma$ have to satisfy the  conditions
$\alpha(u) > 0$ and $\alpha''(u)/\alpha(u) < 0$, so that  the metric has the correct 
signs and the scalar field is real. 

From Eq.(\ref{3.9}) we find that $ u   \sim 
\gamma_{0}\lambda$, where the proper time $\lambda$ was chosen such
that $u = 0$ corresponds to $\lambda = 0$. Then, the distortion,
which is proportional to the twice integrals of $R_{(\alpha)(\beta)}$ 
with respect to the proper time $\lambda$, is given by
\bqn
\lb{3.14}
D_{(\alpha)(\beta)} &\equiv& \int{d\lambda\int{R_{(\alpha)(\beta)} d\lambda}} \nb\\
 &\sim& \cases{ \lambda\left(\ln\lambda -1\right), &
$\gamma = 1$,\cr
\ln\lambda, & $\gamma = 2$,\cr
\lambda^{2-\gamma}, & $\gamma \not= 1, 2$,\cr}
\eqn
as $\lambda \rightarrow 0$. To study these solutions further, let us consider
the following cases separately.

\subsubsection{$\gamma < 0$}

\begin{figure}
\includegraphics[width=\columnwidth]{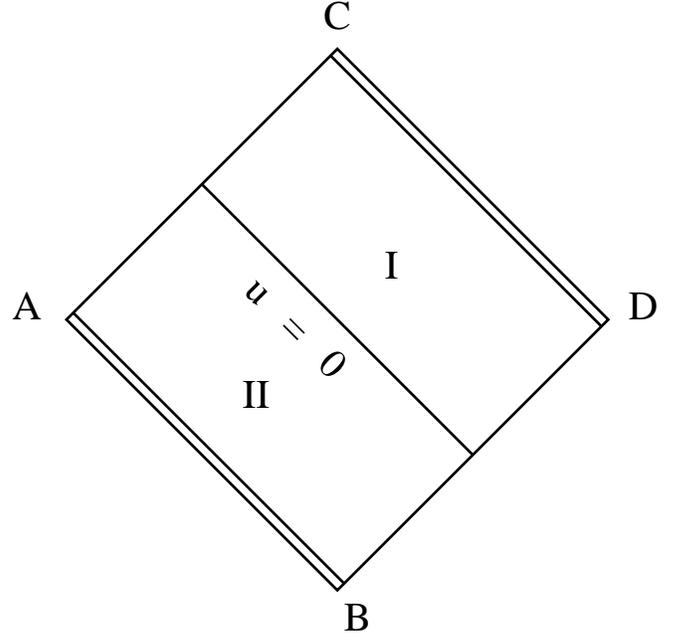}
\caption{The Penrose diagram for the case $\gamma < 0$ of the solution (\ref{3.13})
in the Einstein frame. The double solid lines $AB$ and $CD$ represent null infinities
$u = \pm \infty$ and denote strong spacetime singularities, where both the tidal 
forces and distortions exerting on a freely falling observer become unbound. The spacetime
is regular at $u = 0$, and its extension to Region $II$ where $u < 0$ is unique only 
when $-\gamma$ is an integer. 
}
\label{fig1a}
\end{figure}

In this case,  Eqs.(\ref{3.12}), (\ref{3.13}) and (\ref{3.14}) show that both the tidal forces
and distortions are finite at $u = 0$. Therfore, to have a geodesically maximal spacetime,
we need to extend the solutions across this surface to the region $u < 0$. When $\gamma 
= - n$, where $n (= 1, 2, ...)$ is an integer, the extension is simple, and can be obtained
by simply taking $u$ to be $u \in (-\infty, \infty)$. However, When $- \gamma$ is not an
integer, the solution is not anyalytical at $u = 0$, and the extension is not unique.
One possible extension is to replace $u$ by $|u|$.  Once such an extension is done, it can be
seen that  both the tidal forces and distortions diverge at $|u| \rightarrow \infty$. Therefore,
the spacetimes are singular at the null infinities, and the nature of the singularities is
{\em strong}, as both of them diverge. The corresponding Penrose diagram is given by Fig. \ref{fig1a}.

\subsubsection{$\gamma = 0$}

\begin{figure}
\includegraphics[width=\columnwidth]{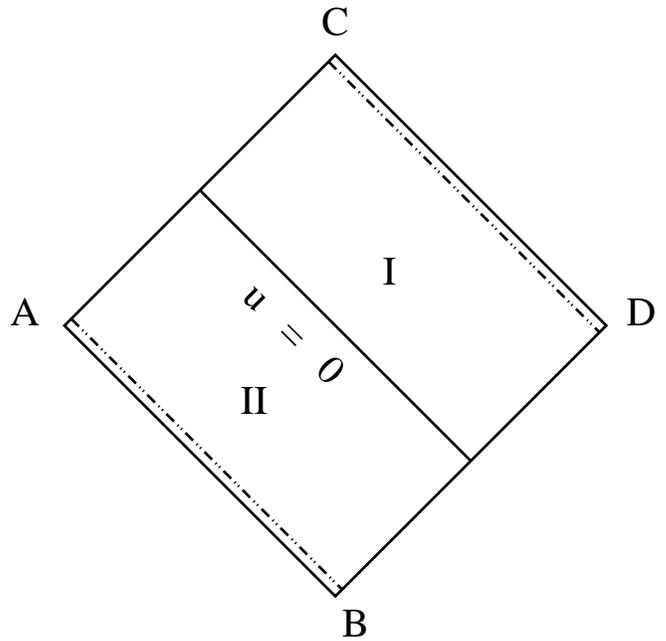}
\caption{The Penrose diagram for the case $\gamma = 0$ of the solution (\ref{3.12b})
in the Einstein frame. The lines $AB$ and $CD$ represent null infinities
$u = \pm \infty$ and the spacetime is singular there, in the sense that
the  distortation becomes unbounded along these null infinities, although 
the tidal forces remain finite.  
}
\label{fig1b}
\end{figure}

In this case,   $\alpha(u)$ and $\phi(u)$ can be given explicitly,  
\bqn
\lb{3.12b}
 \alpha(u) &=& \alpha_{0} \sin\left(\omega u + \Delta\right),  \nb\\
 \phi(u) &=& \pm \sqrt{\frac{D-2}{\kappa^{2}_{D}}} \; \omega u + \phi_{0},
\eqn
where $\Delta$ and $\phi_{0}$ are the integration constants. From the
above we can see that the $(D-2)$-dimensional space ${\cal{S}}$ collapses to
a point at $\omega u + \Delta = n \pi$. This can also be seen from the 
expansion given by  Eq.(\ref{3.7a}), which  now reads,    
\bq
\lb{3.12c}
 \theta_{n} = \omega \left(D-2\right) 
 \frac{\cos\left(\omega u + \Delta\right)}{\sin\left(\omega u + \Delta\right)}.
\eq
Clearly, $\theta_{n}$ diverges at $\omega u + \Delta = n \pi$. However, a closer
investigation of this solution shows that the spacetime is not singular
at these points. For example, one may consider the tidal forces measured 
by observers that move along timelike geodesics not perpendicular to ${\cal{S}}$.
Without loss of generality, let us consider the timelike geodesics, described
by $u = u(\lambda), \; v = v(\lambda), \; x^{2} = x^{2}(\lambda)$ and
$  x^{i} = x^{i}_{0} = Const.$ Then, it can be shown that the timelike geodesical
equation allows the first integration, and yields,
\bqn
\lb{3.12d}
 \dot{u} &=& \gamma_{0}, \;\;\;
 \dot{v} = \frac{1}{2\gamma_{0}}
           \left(\frac{\beta^{2}_{0}}{\alpha^{2}(u)} + 1\right), \nb\\
\dot{x}^{2} &=& \frac{\beta_{0}}{\alpha^{2}(u)},\;\;\;
\dot{x}^{i} = 0, \; (i = 3, 4, ..., D-1),  
\eqn
where $\gamma_{0}$ and $\beta_{0}$ are two integration constants.
From the above we find the following unit vectors,
\bqn
\lb{3.10a}
e^{\mu}_{(0)} &\equiv& \frac{dx^{\mu}}{d\lambda}
=\gamma_{0}\delta^{\mu}_{u} + \frac{1}{2\gamma_{0}}
\left(\frac{\beta^{2}_{0}}{\alpha^{2}(u)} + 1\right)\delta^{\mu}_{v}
+ \frac{\beta_{0}}{\alpha^{2}(u)}\delta^{\mu}_{2},\nb\\
e^{\mu}_{(1)} &=& \gamma_{0}\delta^{\mu}_{u} + \frac{1}{2\gamma_{0}}
\left(\frac{\beta^{2}_{0}}{\alpha^{2}(u)} - 1\right)\delta^{\mu}_{v}
+ \frac{\beta_{0}}{\alpha^{2}(u)}\delta^{\mu}_{2},\nb\\
e^{\mu}_{(2)} &=& \frac{\beta_{0}}{\gamma_{0}\alpha(u)} \delta^{\mu}_{v} 
     + \frac{1}{\alpha(u)} \delta^{\mu}_{2},\nb\\
e^{\mu}_{(i)} &=&   \frac{1}{\alpha(u)} \delta^{\mu}_{i}, 
\; (i = 3, 4, ..., D-1).
\eqn
It can be shown that they satisfy Eq.(\ref{3.11}), that is, they  also
form a freely-falling frame. Then, we find that
\bq
\lb{3.10b}
R_{(\alpha)(\beta)} \equiv R_{\mu\nu}e^{\mu}_{(\alpha)}e^{\nu}_{(\beta)}  
= \omega^{2} \left(D-2\right)e^{u}_{(\alpha)} e^{u}_{(\beta)},
\eq
which is always finite, as can be seen from Eq.(\ref{3.10a}). Therefore, although
the $(D-2)$-dimensional space ${\cal{S}}$ focuses at $\alpha(u) = 0$,  
no spacetime singularities appear at these points, because no trapped compact
surface exists in the present case. The Hawking-Penrose singularity theorems
require the existence of both a compact trapped surface and the focusing of
geodesics \cite{HE73}. The extension across these focusing points can be done
by simply taking $u$ as any real value, i.e.,  $u \in (-\infty, \infty)$.
Once such an extension is made, Eq.(\ref{3.12b}) shows that we may have
spacetime singularities at $u = \pm \infty$.  As Eq.(\ref{3.14}) shows that
indeed the distortation at these null infinities diverge, although the 
tidal forces still remain finite. Then, the corresponding Penrsoe diagram
is given by Fig. \ref{fig1b}.

\subsubsection{$0 < \gamma < 2$}

\begin{figure}
\includegraphics[width=\columnwidth]{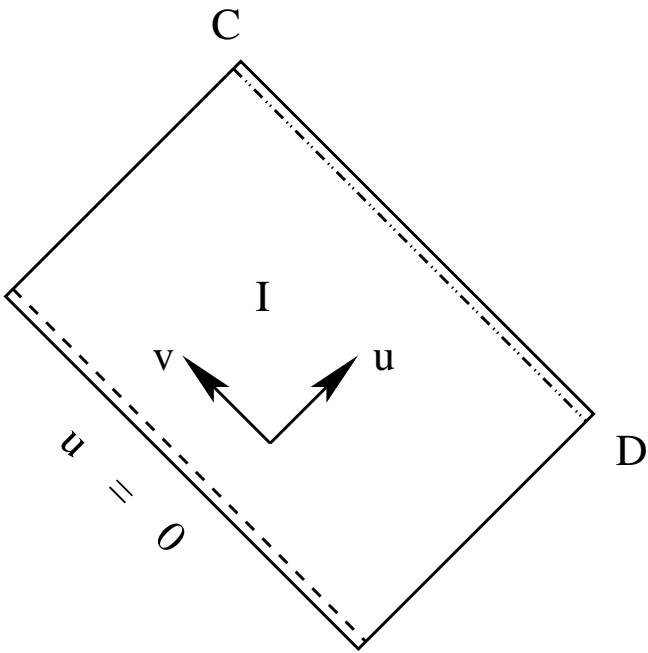}
\caption{The Penrose diagram for the case $0 < \gamma < 2$ of the solution (\ref{3.13}).
The spacetime is singular at $u = 0$ in the sense that the tidal forces become
unbounded, while the distorations remain bounded. 
It is also singular at the null infinity $u = \infty$,  denoted by the line
$CD$, where the tidal forces remain finite, but the distorations become
unbounded.  
}
\label{fig1c}
\end{figure}

In this case, Eqs.(\ref{3.12}), (\ref{3.13}) and (\ref{3.14}) show that  at
$u = 0$ the tidal forces become unbounded, while the distortions remain finite.
This type of   singularities  is usually said   {\em weak}, and the spacetime 
beyond this surface may be extendible \cite{Ori921,Ori922}, although it is still unclear
how to carry out specifically such extensions. As $u \rightarrow \infty$,
from Eqs.(\ref{3.12}), (\ref{3.13}) and (\ref{3.14}) we find that the tidal 
forces are bounded, but now the distortions become unbounded. The corresponding
Penrose diagram is givne by Fig. \ref{fig1c}.

\subsubsection{$ \gamma = 2$}
 
\begin{figure}
\includegraphics[width=\columnwidth]{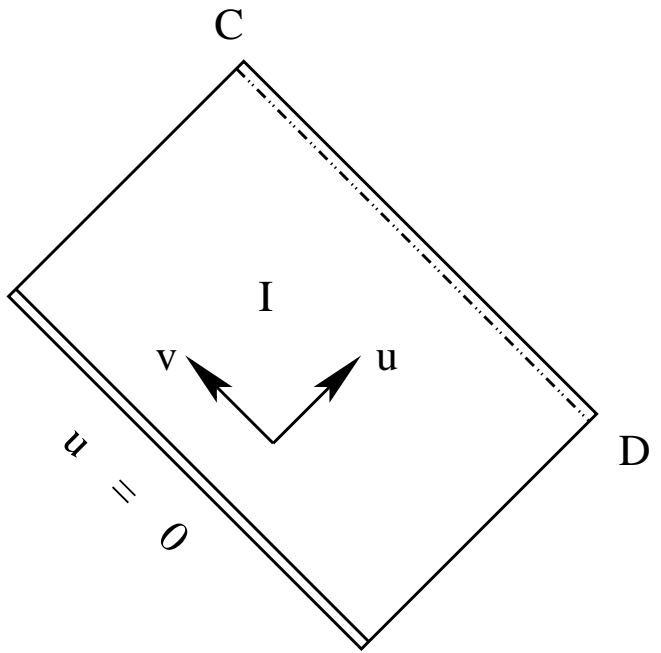}
\caption{The Penrose diagram for the case $  \gamma = 2$ of the solution (\ref{3.13}).
The spacetime is singular at $u = 0$ in the sense that the tidal forces become
unbounded, while the distorations remain bounded. 
It is also singular at the null infinity $u = \infty$,  denoted by the line
$CD$, where the tidal forces remain finite, but the distorations become
unbounded.  
}
\label{fig1d}
\end{figure}

When $\gamma = 2$,   Eq.(\ref{3.13})   has the solution
\bq
\lb{3.16}
\alpha(u) =  \alpha_{0} u^{\delta},  
\eq
where  $\omega^{2} = \delta(1-\delta)$ with $0 < \delta < 1$. For such a
solution, Eq.(\ref{3.7a}) reads,    
\bq
\lb{3.16a}
 \theta_{n} = \frac{\delta \left(D-2\right)}{u}.
\eq
Then, from Eqs.(\ref{3.12}), (\ref{3.13}) and (\ref{3.14}) we find that in 
the present case both  the tidal forces and distortions   become unbound
at $u = 0$, so a strong spacetime singularity appears at $u = 0$. In this
case, we also have $\theta_{n}(u = 0) = \infty$. However, at $u = \infty$,
the tidal forces remain finite, while the distortions   become unbound.
The corresponding Penrose diagram in this case is given by Fig. \ref{fig1d}.

\subsubsection{$\gamma > 2$}

\begin{figure}
\includegraphics[width=\columnwidth]{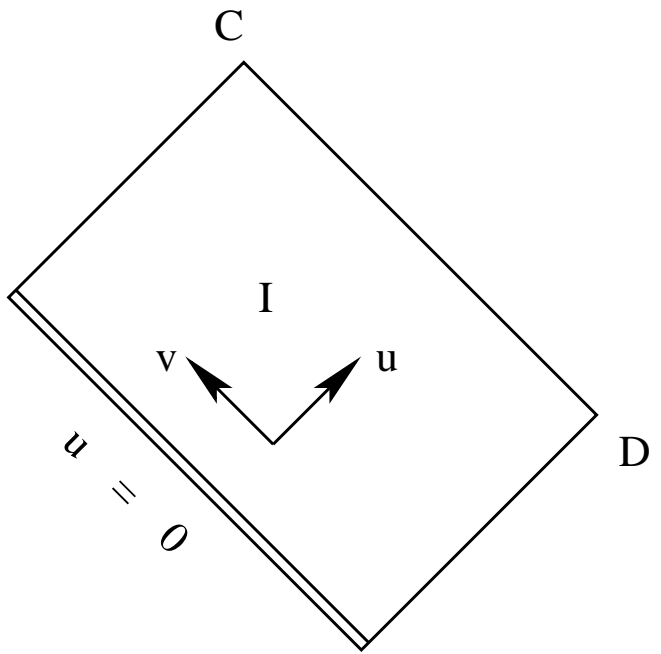}
\caption{The Penrose diagram for the case $  \gamma > 2$ of the solution (\ref{3.13}).
The spacetime is singular at $u = 0$ in the sense that both the tidal forces
and distorations become unbounded. At the null infinity $u = \infty$,  denoted 
by the line $CD$, it is non-singular, because now  both the tidal forces and
distorations  remain finite.  
}
\label{fig1e}
\end{figure}

When $\gamma >  2$, both the  tidal forces and the distortion
become unbound at $u = 0$, so the spacetime has a strong singularity along 
this surface. However, at $u = \infty$ all of them remain finite. Therefore,
in the present case the spacetime is free of spacetime singularity at the 
null infinity $ u = \infty$. The corresponding Penrose diagram in this case 
is given by  Fig. \ref{fig1e}.

\subsection{ $F'(u) G'(v)  \not= 0$}

In this case to solve Eqs.(\ref{3.4a})-(\ref{3.4e}), it is found convenient first to
intruduce two new coordinates $\bar{u}$ and $\bar{v}$ via the relations 
$\bar{u} \equiv F(u)$ and $\bar{v} \equiv G(v)$,  using the gauge freedom (\ref{3.8}).
In terms of these new coordinates, the metric (\ref{3.2}) takes the form,
\bqn
\lb{3.2a}
ds^{2}_{D,E} &=& \bar{g}_{\mu\nu}d\bar{x}^{\mu}d\bar{x}^{\nu} \nb\\
&=& 2 e^{2\Sigma(\bar{u}, \bar{v})}d\bar{u} d\bar{v} 
- e^{2H(\bar{u}, \bar{v})} d\Sigma^{2}_{D-2},
\eqn
where 
\bqn
\lb{3.3a} 
H(\bar{u}, \bar{v}) &\equiv&  h(u,v)  
=  \frac{1}{D-2}\ln\left(\bar{u} + \bar{v}\right),\nb\\
\Sigma(\bar{u}, \bar{v}) &\equiv&     \sigma({u}, {v}) - \frac{1}{2}\ln\left[F'(u)G'(v)\right]. 
\eqn
Then, it can be shown that Eqs.(\ref{3.4a})-(\ref{3.4e}) reduce to
\bqn
\lb{3.16aa}
M_{,t} &=& \frac{1}{2} t\left({\phi_{,t}}^{2} + {\phi_{,y}}^{2}\right),\\
\lb{3.16ab}
M_{,y} &=&   t{\phi_{,t}}{\phi_{,y}},\\
\lb{3.16ac}
M_{,tt} &-& M_{,yy}  = - \frac{1}{2} \left({\phi_{,t}}^{2} - {\phi_{,y}}^{2}\right),\\
\lb{3.16ad}
\phi_{,tt} &+& \frac{1}{t} \phi_{,t} - \phi_{,yy} = 0,
\eqn
where
\bqn
\lb{3.16ae}
\Sigma  &\equiv& - \frac{D-3}{2(D-2)} \ln(t) + \kappa^{2}_{D} M,\nb\\
t &\equiv& \bar{u} + \bar{v},\;\;\; y \equiv \bar{u} - \bar{v}.
\eqn
Eq.(\ref{3.16ac}) is the integrability condition of Eqs.(\ref{3.16aa}) and (\ref{3.16ab}).
Thus, once a solution for $\phi$ is found from Eq.(\ref{3.16ad}), the remaining is to find $M$ 
from Eqs.(\ref{3.16aa}) and (\ref{3.16ab}) by quadratures. 

Using the freedom on the choice of the two null vectors defined by Eq.(\ref{3.8c}), for the
metric (\ref{3.2a}) we define $\bar{l}_{\mu} = \delta^{\bar{u}}_{\mu}$ and
$ \bar{n}_{\mu} = \delta^{\bar{v}}_{\mu}$. Then, we find that,
\bqn
\lb{3.16af}
\bar{\theta}_{l} &\equiv& \bar{g}^{\alpha\beta}\bar{l}_{\alpha;\beta}
= \frac{e^{-2\Sigma}}{t},\nb\\
\bar{\theta}_{n} &\equiv& \bar{g}^{\alpha\beta}\bar{n}_{\alpha;\beta}
= \frac{e^{-2\Sigma}}{t},
\eqn
from which we find that $\bar{\theta}_{l}\bar{\theta}_{n} \ge 0$, where equality holds
only when $t = \infty$ or/and $\Sigma = \infty$. Therefore, in the present case, the
$(D-2)$-dimensional surface ${\cal{S}}$ is always trapped or marginally trapped. However,
it must be noted that the coordinates $(t, y)$ or $(\bar{u}, \bar{v})$ do not always
cover the whole spacetime. There exist cases where the hypersurface $\Sigma = \infty$
does not represents the boundary of the spacetime. To have a geodesically maximal 
spacetime, extension beyond this surface is needed. In the extended region(s), one
may have $\bar{\theta}_{l}\bar{\theta}_{n} < 0$, that is, the spacetime is not trapped.
Typical examples of this kind can be found in \cite{Wang1,Wang2,Wang3,Wang4}. In the following, we consider 
three classes of solutions, to be referred, respectively, to as, Class IIa, IIb, 
and IIc, and show explicitly that such cases happen here, too.  

\subsubsection{Class IIa Solutions}

\begin{figure}
\includegraphics[width=\columnwidth]{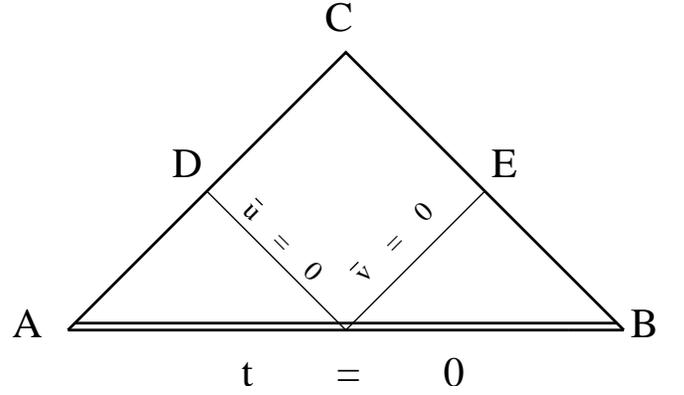}
\caption{The Penrose diagram for the solutions given by Eq.(\ref{3.16a1}) in D-dimensional
spacetime. The horizontal line $t = 0$ represents a big bang  singularity.
The $(D-2)$-dimensional surfaces of constant $\bar{u}$
and $\bar{v}$ are always trapped.}
\label{fig2a}
\end{figure}

This class of solutions is given by
\bqn
\lb{3.16a1}
M    &=& \frac{1}{2} c^{2} \ln(t) + M_{0},\nb\\
\phi &=& c \ln(t) + \phi_{0},
\eqn
where $c,\; \phi_{0}$ and $M_{0}$ are integration constants. Then, from Eq.(\ref{3.16af})
we find that
\bq
\lb{3.16a2}
\bar{\theta}_{l} = \bar{\theta}_{n} = \frac{e^{-\kappa^{2}_{D}M_{0}}}{t^{\chi^{2} +
\frac{1}{D-2}}},
\eq
which means that $t = 0$ represents a focusing point, where 
$\chi \equiv c^{2}\kappa_{D}^{2}$. Since now all the surfaces
${\cal{S}}$ of constant $t$ and $y$ are trapped, then, according to the singularity
theorems \cite{HE73}, the spacetime must be singular at $t = 0$, as can be seen clearly
from the expression,
\bq
\lb{3.16a3}
R_{D}[g] \equiv   \kappa_{D}^{2}g^{\alpha\beta}\phi_{,\alpha}
\phi_{,\beta}= \frac{A_{0}}{t^{\chi^{2} + \frac{D-1}{D-2}}},
\eq
where $A_{0} \equiv 2\chi^{2} e^{-2\kappa^{2}_{D}M_{0}}$.
 The corresponding Penrose diagram is given by 
Fig. \ref{fig2a}.

\subsubsection{Class IIb Solutions}

This class of solutions is given by
\bqn
\lb{3.16a4}
M &=& \frac{1}{2} c^{2} \ln\left(\frac{t^{4}}{\left(y^{2} - t^{2}\right)
\left(y + \sqrt{y^{2} - t^{2}}\right)^{2}}\right) + M_{0},\nb\\
\phi &=& c \ln\left(\frac{t^{2}}{y + \sqrt{y^{2} - t^{2}}}\right) + \phi_{0},
\eqn
for which from Eq.(\ref{3.16a2}) we find that
\bq
\lb{3.16ah}
\bar{\theta}_{l} = \bar{\theta}_{n} = 
\frac{\left\{\left(y^{2} - t^{2}\right)\left[y
+ \sqrt{y^{2} - t^{2}}\right]^{2}\right\}^{\chi^{2}}}
{e^{2\kappa^{2}_{D}M_{0}}t^{4\chi^{2} +
\frac{1}{D-2}}}.
\eq
Clearly, in the present case the hypersurface $t = 0$ still represents a focusing
point. In addition, on the hypersurfaces $y^{2} = t^{2}$, we have $\Sigma = \infty$
and $\bar{\theta}_{l} \bar{\theta}_{n} = 0$. Thus, the surfaces ${\cal{S}}$ now become
marginally trapped along $y^{2} = t^{2}$. If the spacetime is not singular on 
these null surfaces, extension beyond them is needed. To study the singular behavior 
along these surfaces, let us consider the quantity,
\bqn
\lb{3.16ai}
R_{D}[g] &\equiv&   \kappa_{D}^{2}g^{\alpha\beta}\phi_{,\alpha}
\phi_{,\beta}\nb\\
&=&  \frac{2A_{0}\left(4y^{2} - 3 t^{2} + 2\right)}
{\left[\sqrt{y^{2} - t^{2}}\left(y + \sqrt{y^{2} - t^{2}}\right)\right]^{1-2\chi^{2}}}\nb\\
& & \times t^{- \left(4\chi^{2} + \frac{D-1}{D-2}\right)}.
\eqn
Clearly, the spacetime is singular at $t = 0$. The singular behavior along the hypersurfaces
$y^{2} = t^{2}$ depend on the values of $\chi$.

\begin{figure}
\includegraphics[width=\columnwidth]{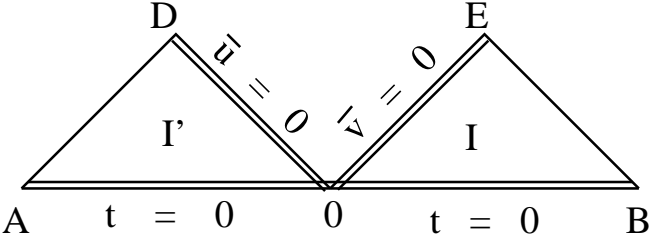}
\caption{The Penrose diagram in the D-dimensional spacetime 
for the solutions given by Eq.(\ref{3.16a4}),
as well as for the ones given by Eq.(\ref{3.16am}),
for $\chi^{2} < 1/2$. The horizontal line $t = 0$ 
represents a big bang  singularity. The $(D-2)$-dimensional surfaces of 
constant $\bar{u}$ and $\bar{v}$ are always trapped in Regions $I$ and $I'$.
The spacetime along the line $0D$ and $0E$ are singular.   The two regions
$I$ and $I'$ are physically disconnected in both cases. 
%
}
\label{fig2b}
\end{figure}

\noindent{\bf Case B.2.1: $\chi^{2} < \frac{1}{2}$}. In this case, 
 it is also singular on the null hypersurfaces $y^{2} =  t^{2}$, and the 
corresponding Penrose diagram is given by Fig. \ref{fig2b}. 

\noindent{\bf Case B.2.2: $\frac{1}{2} \le \chi^{2} < 1$.} In this case,  
the spacetime is not singular at  $y = \pm t$, but the metric
coefficient $\Sigma$ is. To extend the metric beyond these surfaces, 
because of the symmetry, it is suffcient to consider   the extension 
only across the hypersurface $y = t$ from the region $\bar{u} > 0,\; 
\bar{v} < 0$. The extension across the hypersurface $y = -t$ from the region 
$\bar{u} < 0,\; \bar{v} > 0$ is similar, and can be obtained by exchanging the 
roles of $\bar{u}$ and $\bar{v}$. Then, using the gauge freedom 
(\ref{3.8}), we  introduce two new coordinates 
$\tilde{u}$ and $\tilde{v}$  via the relations,
\bq
\lb{3.16aj}
 \bar{u} =  \tilde{u}^{2n},  \;\;\;
 \bar{v} =  - \left(-\tilde{v}\right)^{2n}, 
 \eq
where 
\bq
\lb{3.16ak}
n \equiv \frac{1}{2(1- \chi^{2})}.
\eq
It can be shown that the solutions are analytical across the hypersurfaces $y = \pm t$ 
only when $n$ is an integer. Otherwise, the solutions are not analytical, and the extension
across $y = \pm t$ in not unique. Therefore, in the following we shall consider only
the case where $n$ is an integer. Then, in terms of $\tilde{u}$ and $\tilde{v}$, the
solutions read
\bqn
\lb{3.2aa}
ds^{2}_{D,E} &=& \tilde{g}_{\mu\nu}d\tilde{x}^{\mu}d\tilde{x}^{\nu} \nb\\
&=& 2 e^{2\tilde{\Sigma}} d\tilde{u} d\tilde{v} 
- e^{2\tilde{H}} d\Sigma^{2}_{D-2},
\eqn
where 
\bqn
\lb{3.3a1} 
\tilde{H} &=&    \frac{1}{D-2}\ln\left(\tilde{u}^{2n} - \left(-\tilde{v}\right)^{2n}\right),\nb\\
\tilde{\Sigma} &=&    \left(2\chi^{2} - \frac{D-3}{2(D-2)}\right)
                     \ln\left(\tilde{u}^{n} - \left(-\tilde{v}\right)^{n}\right)\nb\\
	& & - \frac{D-3}{2(D-2)} \ln\left(\tilde{u}^{n} + \left(-\tilde{v}\right)^{n}\right)
	    + \tilde{\Sigma}_{0},\nb\\
\phi &=&  \frac{2\chi}{\kappa_{D}} 
                     \ln\left(\tilde{u}^{n} - \left(-\tilde{v}\right)^{n}\right)
		     + \phi_{0}, 
\eqn
where $\tilde{\Sigma}_{0} \equiv \kappa^{2}_{D}M_{0} + \ln\left(2^{3/2 - \chi^{2}} n\right)$.
Clearly, the coordinate singularity at $y = t$ or $\tilde{v} = 0$ disappears, and the solutions 
can be considered as   valid for $\tilde{v} > 0$.  To study the properties of the spacetime
in the extended region, let us consider the quantities, 
\bqn
\lb{3.3aa} 
\tilde{\theta}_{l} &=& \left(D-2\right)e^{-2\tilde{\Sigma}}\tilde{H}_{,\tilde{v}}
= 2n e^{-2\tilde{\Sigma}}\frac{\left(-\tilde{v}\right)^{2n-1}}{\tilde{u}^{2n}
             - \left(-\tilde{v}\right)^{2n}},\nb\\
\tilde{\theta}_{n} &=& \left(D-2\right)e^{-2\tilde{\Sigma}}\tilde{H}_{,\tilde{u}}
= 2n e^{-2\tilde{\Sigma}}\frac{\tilde{u}^{2n-1}}{\tilde{u}^{2n}
             - \left(-\tilde{v}\right)^{2n}},\nb\\
{\cal{L}}_{n}\tilde{\theta}_{l}	 &=& - 4n^{2}e^{-4\tilde{\Sigma}}
\frac{\tilde{u}^{n-1}\left(-\tilde{v}\right)^{2n-1}}{\left(\tilde{u}^{2n}
             - \left(-\tilde{v}\right)^{2n}\right)^{2}}\nb\\
& & \times\left\{\left(2\chi^{2} + \frac{1}{D-2}\right)\tilde{u}^{n} 
    + 2\chi^{2}\left(-\tilde{v}\right)^{n}\right\},\nb\\    
R_{D}\left[\tilde{g}\right] &=& 8 n^{2}\chi^{2}e^{-2\tilde{\Sigma}_{0}}\nb\\
& & \times \frac{\left(\tilde{u}^{n} + \left(-\tilde{v}\right)^{n}\right)^{\frac{D-3}{D-2}}
\left(-\tilde{u}\tilde{v}\right)^{n-1}}
{\left(\tilde{u}^{n} - \left(-\tilde{v}\right)^{n}\right)^{4\chi^{2} + \frac{D-1}{D-2}}}.
\eqn
Then, we find that
\bqn
\lb{3.3ab}
\tilde{\theta}_{l} \tilde{\theta}_{n} &=& 4n^{2}e^{-4\tilde{\Sigma}}
\frac{\left(-\tilde{u} \tilde{v}\right)^{2n-1}}
{\left(\tilde{u}^{2n} - \left(-\tilde{v}\right)^{2n}\right)^{2}}\nb\\
&=&\cases{< 0, \tilde{u} \tilde{v} >0,\cr
= 0, \tilde{u} \tilde{v} = 0,\cr
> 0, \tilde{u} \tilde{v} < 0.\cr}
\eqn
That is, now in the extended region where $\tilde{u} \tilde{v} >0$ the  spacetime
becomes untrapped. Across the hypersurface $\tilde{v} = 0,\; \tilde{u} \ge 0$,
we have
\bqn
\lb{3.3ac}
& & \tilde{\theta}_{l}\left(\tilde{u}>0, \tilde{v} = 0\right) = 0,\;\;\;
\tilde{\theta}_{n}\left(\tilde{u}>0, \tilde{v} =0 \right)   >0,\nb\\
& & {\cal{L}}_{n}\tilde{\theta}_{l} \left(\tilde{u}> 0, \tilde{v} = 0\right) = 0.
\eqn
Therefore, the half infinite line $\tilde{v} = 0,\; \tilde{u} \ge 0$ represents
a past degenerate apparent horizon. 

\begin{figure}
\includegraphics[width=\columnwidth]{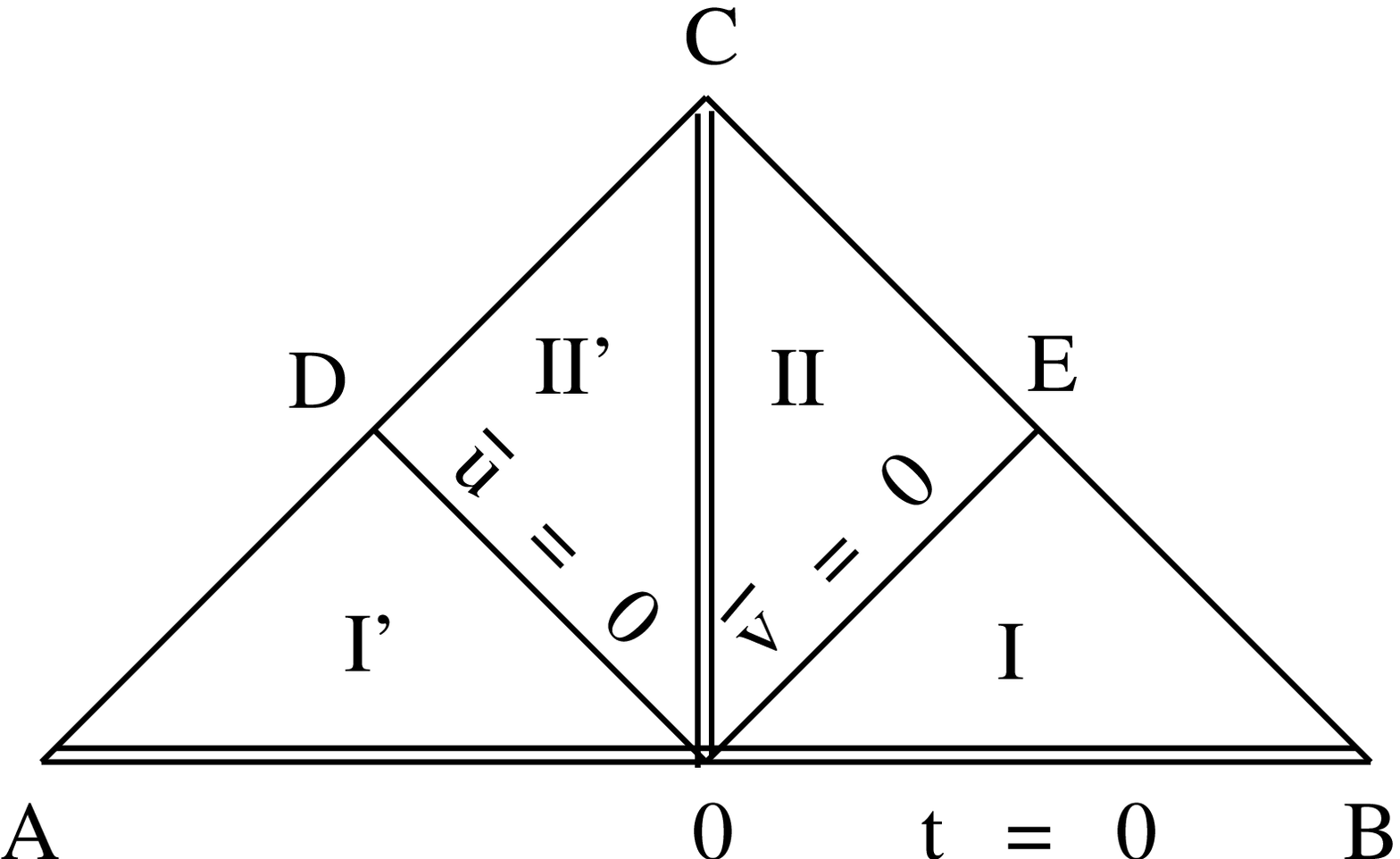}
\caption{The Penrose diagram in the D-dimensional spacetime for $\frac{1}{2} \le 
\chi^{2} < 1$ for the solutions given by Eq.(\ref{3.2aa}) with
$n$ being an even integer, as well as for the ones  given by Eq.(\ref{3.16am}) with
 $n$ being an odd integer. The horizontal line $t = 0$ 
represents a big bang  singularity. The $(D-2)$-dimensional surfaces of 
constant $\bar{u}$ and $\bar{v}$ are always trapped in Regions $I$ and $I'$,
but not in Regions $II$ and $II'$. The spacetime along the vertical
line $0C$ is singular.   The two regions
$II$ and $II'$ are physically disconnected. The lines $0E$ and $0D$ represent
 past degenerate apparent horizons. 
%
}
\label{fig2c}
\end{figure}

To study the singular behavior of the spacetime in the extended region,
we need to distinguish the case where $n$ is an even interger from the one where $n$
is an odd integer.  When $n$ is an even interger, Eq.(\ref{3.3aa}) shows that the spacetime
is singular along the line $\tilde{u} = \tilde{v}$, denoted by the vertical line
$0C$ in Fig. \ref{fig2c}. 

Because of the symmetry of the spacetime, one can make
a similar extension across the half line $\bar{u} = 0$ and $\bar{v} \ge 0$, but
now with
\bq
\lb{3.16ak2}
 \bar{u} =  - \left(-\tilde{u}\right)^{2n},  \;\;\;
 \bar{v} =   \tilde{v}^{2n}.
 \eq
Then, one finds that  this half line also  represents a past degenerate apparent
horizon, and the extended region $II'$ is not trapped, although it is disconnected
with Region $II$, because of the timelike singularity along the vertical line
$0C$. 

\begin{figure}
\includegraphics[width=\columnwidth]{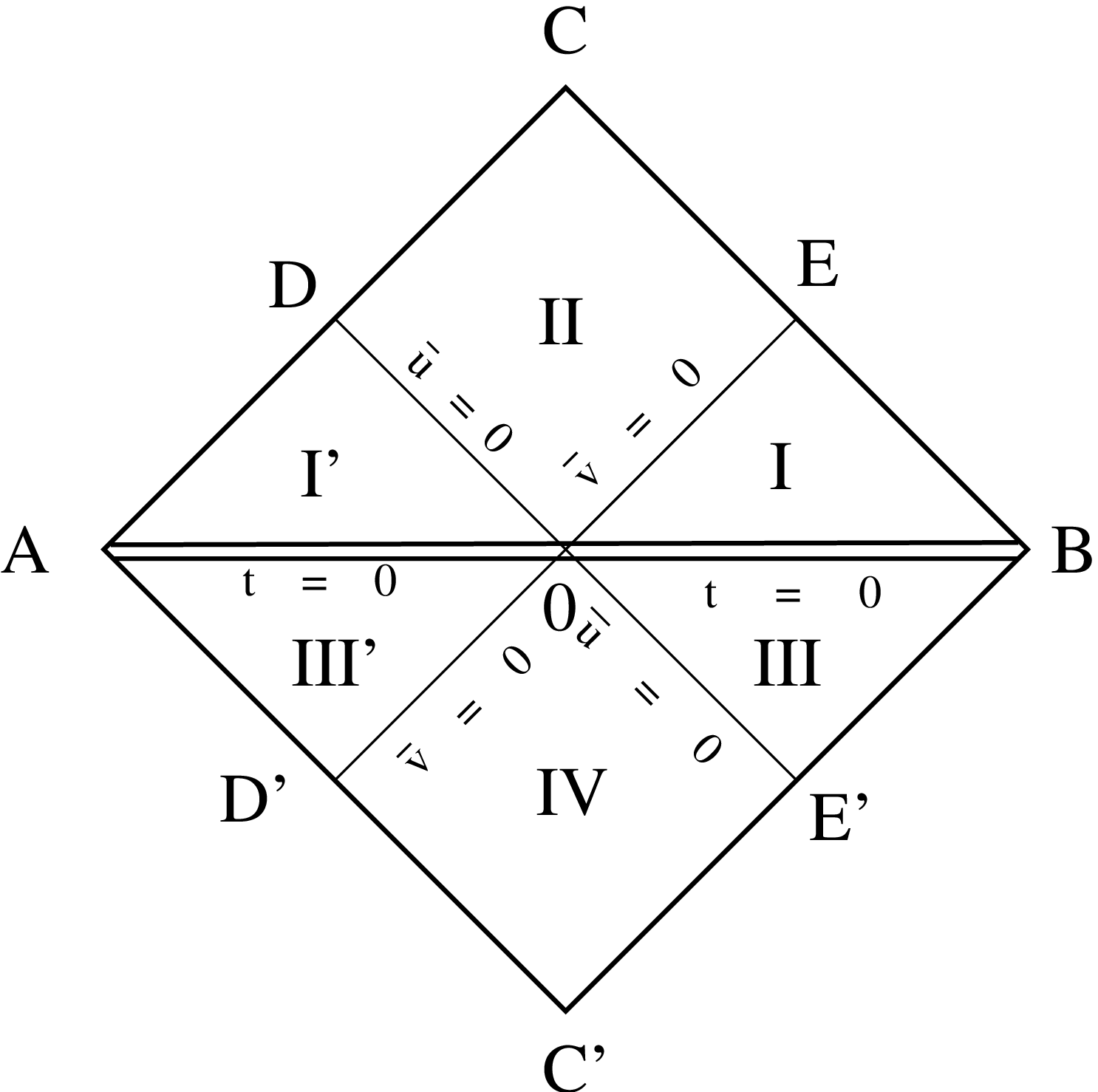}
\caption{The Penrose diagram in the D-dimensional spacetime for $\frac{1}{2} \le 
\chi^{2} < 1$ for the solutions given by Eq.(\ref{3.2aa}) with
$n$ being an odd integer, as well as for the ones  given by Eq.(\ref{3.16am}) with
 $n$ being an even integer. The horizontal line $t = 0$ 
represents a spacetime  singularity. The $(D-2)$-dimensional surfaces of 
constant $\tilde{u}$ and $\tilde{v}$ are always trapped in Regions $I$,
$I'$, $III$ and $III'$, but not in Regions $II$, and $IV$.   The upper
half part is disconnected with the lower half part by the spacetime
singularity along the line $AB$ where $t=0$. The lines $0E$ and $0D$ represent
 past degenerate apparent horizons, while the ones   $0E'$ and $0D'$ represent
 future degenerate apparent horizons. Regions $I$ and $I'$ represent white holes,
 while Regions $III$ and $III'$ represent black holes.
%
}
\label{fig2d}
\end{figure}

When $n$ is an odd interger, Eq.(\ref{3.3aa}) shows that the spacetime
is not singular along the line $\tilde{u} = \tilde{v}$, as shown in Fig. \ref{fig2d},
where the lines $0E$ and $0D$ represent past degenerate apparent horizons.
Then, Regions $I$ and $I'$ act as white holes. Since the solutions are symmetric
with respect to $t$, one may consider the case where $t \le 0$. Then, the spacetime
will be given by the lower half part of Fig. \ref{fig2d}, in which the $(D-2)$-dimensional
surfaces ${\cal{S}}$ of constant $\tilde{u}$ and $\tilde{v}$ are trapped in
Regions $III$ and $III'$, and is not trapped in   Region $IV$. The lines $0D'$
and $0E'$ act as future degenerate apparent horizons, so  Regions $III$ and $III'$ 
now represent black holes. the spacetimes of the lower half part is disconnected with 
that of the upper half by the spacelike singularity located along $t = 0$.  

\noindent{\bf Case B.2.3: $\chi^{2} \ge 1$.}
When $\chi^{2} \ge 1$, the hypersurfaces $y = \pm t$  represent spacetime null infinities, 
and the solutions are already geodesically maximal. Indeed, it is found that the null goedesics 
$\bar{u} = $ Constant have the integral,
\bq
\lb{3.16al}
\eta = \cases{\eta_{0} \left(-\bar{v}\right)^{1- \chi^{2}}, & $\chi^{2} > 1$,\cr
\eta_{0} \ln(- \bar{v}), & $\chi^{2} = 1$,\cr}
\eq
near the hypersurface $ y = t\; (\bar{v} = 0)$, where $\eta_{0}$ is an integration constant, and
$\eta$ denotes the affine parameter along the null geodesics.
Thus, as $\bar{v} \rightarrow 0^{-}$, we always have $|\eta| \rightarrow \infty$. 

\begin{figure}
\includegraphics[width=\columnwidth]{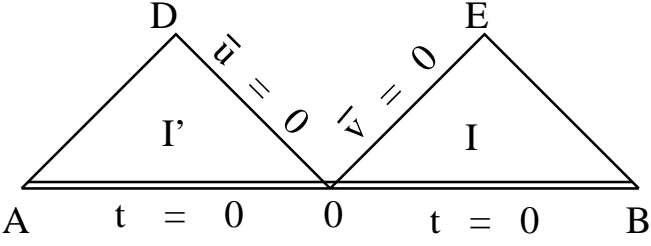}
\caption{The Penrose diagram for the solutions given by Eq.(\ref{3.16a4}) and
for the ones  given by Eq.(\ref{3.16am})
in the D-dimensional spacetime for $\chi^{2} \ge 1$. The horizontal line $t = 0$ 
represents a big bang  singularity. The $(D-2)$-dimensional surfaces of 
constant $\bar{u}$ and $\bar{v}$ are always trapped in Regions $I$ and $I'$,
and are physically disconnected in both cases. The  line $0D$ and $0E$ represent 
spacetime null infinities, and are not singular.   
%
}
\label{fig2e}
\end{figure}

Similar, it can be shown that the hypersurface $y = -t$ also represents a null infinity of the 
spacetime. Therefore, the 
corresponding Penrose diagram  is given by  Fig. \ref{fig2e}. 

\subsubsection{Class IIc Solutions}

This class of solutions is given by
\bqn
\lb{3.16am}
M &=& \frac{1}{2} c^{2} \ln\left(\frac{\left(y + \sqrt{y^{2} - t^{2}}\right)^{2}}
{y^{2} - t^{2}}\right) + M_{0},\nb\\
\phi &=& c \ln\left(y + \sqrt{y^{2} - t^{2}}\right) + \phi_{0},
\eqn
for which we find
\bqn
\lb{3.16ana}
\bar{\theta}_{l} &=& \bar{\theta}_{n} = \frac{e^{-2\kappa^{2}_{D}M_{0}}
         \left(y^{2} - t^{2}\right)^{\chi^{2}}}
	 {t^{\frac{1}{D-2}} \left(y + \sqrt{y^{2} - t^{2}}\right)^{2\chi^{2}}}\nb\\
&\propto& \cases{ t^{-\frac{1}{D-2}}, & $y > 0$,\cr
t^{-\left(4\chi^{2} + \frac{1}{D-2}\right)}, & $ y < 0$,\cr}
\eqn
as $ t \rightarrow 0$. Therefore, in this case the hypersurface $t =0$ represents a 
focusing point, while the ones $y = \pm t$ may represent some kind of horizons. 
To see their exact natures, let us consider the quantity,
\bqn
\lb{3.16an}
R_{D}[g] &=&   \kappa_{D}^{2}g^{\alpha\beta}\phi_{,\alpha}
\phi_{,\beta}   \nb\\
&=& - \frac{2A_{0}\left(y^{2} -  t^{2}\right)^{\chi^{2} -1/2}}
{\left(y + \sqrt{y^{2} - t^{2}}\right)^{2\chi^{2}+1}}
\; t^{\frac{D-3}{D-2}}.
\eqn
Thus, the nature of the spacetime singularity at $t = 0$ is different on the half line
$y > 0$  from that on the other half line  $y < 0$. In particular, we find
\bqn
\lb{3.16anb}
R_{D}[g]  &\propto& \cases{ - t^{\frac{D-3}{D-2}} \rightarrow 0, & $y > 0$,\cr
t^{-\left(4\chi^{2} + \frac{1}{D-2}\right)} \rightarrow - \infty, & $ y < 0$,\cr} 
\eqn
as  $ t \rightarrow 0$. Therefore, $R_{D}[g]$ is singular only on the half
line $t = 0, \; y \le 0$. However, the studies of other scalars show that the
other half line, $t = 0$ and $y > 0$, is also singular. For example, the 
corresponding Kretschmann scalar is given by
\bqn
\lb{Kretsch}
I_{D} &\equiv& R^{\mu\nu\lambda\sigma}R_{\mu\nu\lambda\sigma}\nb\\
&=&\left. \frac{\left(y^{2} - t^{2}\right)^{2\chi^{2} -5}}{8 t^{3}
\left(y + \sqrt{y^{2} - t^{2}}\right)^{4\chi^{2}}}I^{(0)}_{D}(t,y)\right|_{D=4},
\eqn
where, as $t \rightarrow 0$, we have
\bq
\lb{Kretsch2}
I^{(0)}_{4}  \simeq \cases{- 9y^{10}, & $y > 0$,\cr
- \left(1024\chi^{6} -320\chi^{4} + 96 \chi^{2} + 9\right)y^{10}, & $y < 0$.\cr}
\eq

On the other hand, depending on the value of $\chi$, the spacetime may or may not
be singular on the  two null surfaces $y = \pm t$, similar to the last case.
In particular,  when $\chi^{2} < 1/2$, the spacetime is singular there,   
and the the corresponding Penrose diagram is also given  by Fig. \ref{fig2b}.

When $ \frac{1}{2} \le \chi^{2} < 1$,  the spacetime is not singular at  
$y = \pm t$, although the metric coefficients are. Then, extending the solutions
to the region $\tilde{v} > 0$ is needed. As in the last case, the extension
across this surface is uniqe only when $n$ defined by Eq.(\ref{3.16ak}) is an
integer. Thus, in the following we shall consider only this case. Then, the 
extension can  be done by  introducing the new  coordinates  $\tilde{u}$ and
$\tilde{v}$ defined by Eq.(\ref{3.16aj}), for which we find the extended metric 
takes the same form as that given by Eq.(\ref{3.2aa}), but now with 
\bqn
\lb{3.3a1a} 
\tilde{H} &=&    \frac{1}{D-2}\ln\left(\tilde{u}^{2n} - \tilde{v}^{2n}\right),\nb\\
\tilde{\Sigma} &=&    \left(2\chi^{2} - \frac{D-3}{2(D-2)}\right)
                     \ln\left(\tilde{u}^{n} + \left(-\tilde{v}\right)^{n}\right)\nb\\
	& & - \frac{D-3}{2(D-2)} \ln\left(\tilde{u}^{n} - \left(-\tilde{v}\right)^{n}\right)
	    + \tilde{\Sigma}_{0},\nb\\
\tilde{\Sigma}_{0} &\equiv& \kappa^{2}_{D}M_{0} 
  + \frac{1}{2}\ln\left(4^{1 - \chi^{2}} n^{2}\right),\nb\\
  \phi &=&  2c    \ln\left(\tilde{u}^{n} + \left(-\tilde{v}\right)^{n}\right)
		     + \phi_{0}.		     
\eqn 
Thus, the coordinate singularity at  $\tilde{v} = 0$ disappears, and the solutions 
can be considered as   valid for $\tilde{v} > 0$, too.  In terms of $\tilde{u}$ and 
$\tilde{v}$, we find   
\bqn
\lb{3.3aa2} 
\tilde{\theta}_{l} &=& \left(D-2\right)e^{-2\tilde{\Sigma}}\tilde{H}_{,\tilde{v}}
= 2n e^{-2\tilde{\Sigma}}\frac{\left(-\tilde{v}\right)^{2n-1}}{\tilde{u}^{2n}
             - \tilde{v}^{2n}},\nb\\
\tilde{\theta}_{n} &=& \left(D-2\right)e^{-2\tilde{\Sigma}}\tilde{H}_{,\tilde{u}}
= 2n e^{-2\tilde{\Sigma}}\frac{\tilde{u}^{2n-1}}{\tilde{u}^{2n}
             - \tilde{v}^{2n}},\nb\\
{\cal{L}}_{n}\tilde{\theta}_{l}	 &=& - 4n^{2}e^{-4\tilde{\Sigma}}
\frac{\tilde{u}^{n-1}\left(-\tilde{v}\right)^{2n-1}}{\left(\tilde{u}^{2n}
             - \tilde{v}^{2n}\right)^{2}}\nb\\
& & \times\left\{\left(2\chi^{2} + \frac{1}{D-2}\right)\tilde{u}^{n} 
    - 2\chi^{2}\left(-\tilde{v}\right)^{n}\right\},\nb\\    
R_{D}\left[\tilde{g}\right] &=& - 4^{\chi^{2} + \frac{1}{2}}
\chi^{2}  e^{-2\kappa^{2}_{D}M_{0}} \nb\\
& & \times \frac{\left(\tilde{u}^{n} - \left(-\tilde{v}\right)^{n}\right)^{\frac{D-3}{D-2}}
\left(-\tilde{u}\tilde{v}\right)^{n-1}}
{\left(\tilde{u}^{n} + \left(-\tilde{v}\right)^{n}\right)^{4\chi^{2} + \frac{D-1}{D-2}}}.
\eqn
From the above expressions, we can see that  in the extended region where 
$\tilde{u} \tilde{v} >0$ the  spacetime
becomes untrapped. Across the hypersurface $\tilde{v} = 0,\; \tilde{u} \ge 0$,
we have $\tilde{\theta}_{l} = 0,\; \tilde{\theta}_{n}   >0$ and 
$ {\cal{L}}_{n}\tilde{\theta}_{l}   = 0$. That is, the half infinite line 
$\tilde{v} = 0,\; \tilde{u} \ge 0$ acts as a past degenerate apparent horizon. 

In addition, in contrast to the last case, the expression of $R_{D}\left[\tilde{g}\right]$
given above tells us that the spacetime is singular along the vertical line
$\tilde{u} = \tilde{v}$ for $n$ being an odd integer, and not singular for 
$n$ being an even integer. The corresponding Penrose diagram is given, respectively, by
Figs. \ref{fig2b} and \ref{fig2c}.

When $\chi^{2} \ge 1$, the hypersurfaces $y = \pm t$ already represent the null 
infinities  of the spacetime, 
and the corresponding Penrose diagram  is that 
of Fig. \ref{fig2e}, where the hypersurrfaces $\bar{u} = 0$ and  
$\bar{v} = 0$  are not singular, and
the two regions, $II$ and $II'$, are still not connected.


\section{Solutions  in ($D+d$)-dimensional Spacetimes in the string Frame} 

\renewcommand{\theequation}{4.\arabic{equation}}
\setcounter{equation}{0}

In ($D+d$)-dimensions, the metric in the string frame takes the form of Eq. (\ref{2.2}). Considering
Eq. (\ref{3.2}), we find that it can be written in the form,
\bqn
\lb{8.1}
d\hat{s}^{2}_{D+d} &=& \hat{g}_{AB}dx^{A} dx^{B} = \gamma_{\mu\nu}dx^{\mu}dx^{\nu} + \hat{\Phi}\hat{\gamma}_{ab}dz^{a}dz^{b}\nb\\
&=& 2e^{2\hat{\sigma}(\hat{u}, \hat{v})}d\hat{u}d\hat{v} - e^{2\hat{h}(\hat{u}, \hat{v})}d\Sigma_{D-2}^{2} \nb\\
& & 
- e^{2\hat{g}(\hat{u}, \hat{v})}\hat{\gamma}_{ab}dz^{a}dz^{b},
\eqn
where
\bqn
\lb{3.17a}
\gamma_{\mu\nu} &=& \exp\left\{\epsilon_{a}
\left(\frac{4\kappa^{2}_{D}d}{(D-2)(D+d-2)}\right)^{1/2}\phi\right\}
 g_{\mu\nu},\nb\\
\hat{\Phi}  &=& e^{\hat{g}} =  \exp\left\{- \epsilon
\left(\frac{(D-2)\kappa^{2}_{D}}{(D+d-2)d}\right)^{1/2}\phi\right\},
\eqn
 as can be seen from Eqs.(\ref{2.7}) and (\ref{2.9}), where $\epsilon_{a} = \pm 1$.
 
 Introducing the two null vectors, $\hat{l}_{A}$ and $\hat{n}_{A}$ via
 \bq
 \lb{8.2}
 \hat{l}_{A} = \delta^{\hat{u}}_{A}, \;\;\;  \hat{n}_{A} = \delta^{\hat{v}}_{A},
 \eq
 once can show that they also define two null affinely defined geodesic congruences,
 $ \hat{l}_{A; B} \hat{l}^{B} =  0 =  \hat{n}_{A; B} \hat{n}^{B}$. The corresponding expansions
 are given by
 \bqn
 \lb{8.3}
 \hat{\theta}_{l} &\equiv&  \hat{l}_{A; B}\hat{g}^{AB}
    =e^{-2\hat{\sigma}}\Big[\big(D-2\big)\hat{h}_{,\hat{v}} + d \hat{g}_{,\hat{v}}\Big],\nb\\
  \hat{\theta}_{n} &\equiv&  \hat{n}_{A; B}\hat{g}^{AB}
   = e^{-2\hat{\sigma}}\Big[\big(D-2\big)\hat{h}_{,\hat{u}} + d \hat{g}_{,\hat{u}}\Big],
 \eqn
 from which we find
  \bqn
 \lb{8.4}
{\cal{L}}_{\hat{n}}  \hat{\theta}_{l} &=&   e^{-4\hat{\sigma}}\Bigg\{\big(D-2\big)\hat{h}_{,\hat{u}\hat{v}} + d \hat{g}_{,\hat{u}\hat{v}}\nb\\
& & ~~~~~~~~
 - 2\hat{\sigma}_{,\hat{u}}\Big[\big(D-2\big)\hat{h}_{,\hat{v}} + d \hat{g}_{,\hat{v}} \Big]\Bigg\},\nb\\
{\cal{L}}_{\hat{l}}  \hat{\theta}_{n} &=&   e^{-4\hat{\sigma}}\Bigg\{\big(D-2\big)\hat{h}_{,\hat{u}\hat{v}} + d \hat{g}_{,\hat{u}\hat{v}}\nb\\
& & ~~~~~~~~
 - 2\hat{\sigma}_{,\hat{v}}\Big[\big(D-2\big)\hat{h}_{,\hat{u}} + d \hat{g}_{,\hat{u}} \Big]\Bigg\}.
  \eqn

\subsection{$F'(u) \not= 0,\; G'(v)  = 0$}

In this case, the modulus  $\phi(u,v)$ is given by Eq.(\ref{3.15}). 
In the $(D + d)$-dimensional 
spacetime, the metric can be written in the form
\bqn
\lb{3.30}
d{\hat{s}}^{2}_{D+d} &=&  2d\hat{u}d\hat{v} 
- e^{2\hat{h}(\hat{u})}d^{2}\Sigma_{D-2}\nb\\
& & \;
  - \hat{\Phi}^{2}(\hat{u}) \hat{\gamma}_{ab}(z) dz^{a} dz^{b},
\eqn
where 
\bqn
\lb{3.30a}
d\hat{u}  &\equiv& e^{2\hat{\sigma}}du,\nb\\
\hat{\sigma} &\equiv& \sigma - \frac{d}{D-2}\ln\hat{\Phi},\nb\\
\hat{h} &\equiv& h - \frac{d}{D-2}\ln\hat{\Phi}.
\eqn
Following what we did in the Einstein frame, we can construct
a free-falling frame in the $(D+d)$-dimensions, given by 
\bqn
\lb{3.34}
e^{A}_{(0)} &=& \hat{\gamma}_{0}\delta^{A}_{\hat{u}} 
+ \frac{1}{2\hat{\gamma}_{0}}\delta^{A}_{v},\nb\\
e^{A}_{(1)} &=& \hat{\gamma}_{0}\delta^{A}_{\hat{u}} 
- \frac{1}{2\hat{\gamma}_{0}}\delta^{A}_{v},\nb\\
e^{A}_{(i)} &=& e^{-\hat{h}}  \delta^{A}_{i},\nb\\
e^{A}_{(b)} &=& \hat{\Phi}^{-1}  \delta^{A}_{b}, 
\eqn
 which satisfies the relations,
\bq
\lb{3.35}
e^{A}_{(C)}e^{B}_{(D)} \hat{g}_{AB} = \eta_{CD},\;\;\;
e^{A}_{(C); B} e^{B}_{(0)} = 0,
\eq
where $\hat{\gamma}_{0}$ is another integration constant.
Then, it can be shown that the Riemann tensor in this case has only two 
independent components, given by
\bqn
\lb{3.33}
\hat{R}^{(i)}_{\;\;\;\; (0)(j)(0)} &=& 
- \hat{\gamma}_{0}^{2}\left(\hat{h}_{,\hat{u}\hat{u}}
+  \hat{h}_{,\hat{u}}^{2}   \right)\delta^{i}_{j},\nb\\
\hat{R}^{(a)}_{ \;\;\;\; (0)(b)(0)} &=& 
- \hat{\gamma}_{0}^{2}\left(\frac{\hat{\Phi}_{,\hat{u}\hat{u}}}
{\hat{\Phi}}\right)\delta^{a}_{b}. 
\eqn

In addition, Eqs.(\ref{8.3}) and (\ref{8.4}) now read,
\bqn
\lb{3.33aa}
\hat{\theta}_{n} &=& \big(D-2\big) \alpha'(u)\exp\Big(-a u^{(2-\gamma)/2}\Big), \nb\\
\hat{\theta}_{l} &=& 0, \;\;\; {\cal{L}}_{\hat{n}}  \hat{\theta}_{l} = 0 = {\cal{L}}_{\hat{n}}  \hat{\theta}_{l}.
\eqn
Then, according the previous
definition, the $(D+d-2)$-surfaxe ${\cal{S}}$ is   always marginally trapped and degenerate.

To study the solutions further, it is found convenient to
consider the two cases $\gamma \not= 2$ and $\gamma =2$ separately.

\subsubsection{$\gamma \not= 2$} 

In this case, we have  
\bqn
\lb{3.31}
d{\hat{u}}  &=& e^{au^{1-\gamma/2}}du,\nb\\
\hat{h}   &=& \frac{1}{2}a u^{1-\gamma/2} + \alpha(u),\nb\\
\hat{\Phi}   &=& e^{b u^{1-\gamma/2}}, 
\eqn
where
\bqn
\lb{3.31a}
a &\equiv& \frac{2\epsilon_{a}}{2-\gamma} 
\left(\frac{4\omega^{2}d}{D+d-2}\right)^{1/2},\nb\\
b &\equiv& - \frac{2\epsilon_{a}}{2-\gamma} 
\left(\frac{\omega^{2}(D-2)^{2}}{(D+d-2)d}\right)^{1/2}.
\eqn
Then, the non-vanishing frame components of the Riemann tensor are given by
\bqn
\lb{3.33a}
\hat{R}^{(i)}_{\;\;\;\; (0)(j)(0)} &=&  
\hat{\gamma}_{0}^{2} e^{-2au^{1-\gamma/2}}\nb\\
& & \times \left\{
\frac{a\gamma(2-\gamma)}{8 u^{\gamma/2+1}}
+ \frac{a^{2} (2-\gamma)^{2} + 16\omega^{2}}{16 u^{\gamma}}\right\}
\delta^{i}_{j},\nb\\
\hat{R}^{(a)}_{\;\;\;\; (0)(b)(0)} &=& 
\hat{\gamma}_{0}^{2} \frac{b(2-\gamma)}{4}e^{-2au^{1-\gamma/2}}\nb\\
& & \times \left\{
\frac{ \gamma }{u^{\gamma/2+1}}
- \frac{(b-a) (2-\gamma)}{u^{\gamma}}\right\}
\delta^{a}_{b}.
\eqn
Therefore, for the choice $\epsilon_{a} = +1$ we have
\bqn
\lb{3.36}
\hat{R}^{(A)}_{\;\;\;\;(0)(B)(0)} &=& \cases{
\infty, & $\gamma> 0$,\cr
{\mbox{constant}}, & $\gamma = 0$,\cr
\infty, & $- 2 < \gamma < 0$,\cr
{\mbox{finite}}, & $\gamma \le -2$,\cr}\\
\lb{3.36a}
\hat{\theta}_{n} &=& \cases{
\infty, & $\gamma> 2$,\cr
{(D-2)\alpha'(0)}, & $\gamma \le 2$,\cr}
\eqn
as $u \rightarrow 0$, but now with $A, B = i, a$. Thus, the
tidal forces  experiencing by a free-falling observer remain finite
in the string frame at $u = 0$ for all the cases, except for the ones 
where $0 < \gamma$ or $ -2 < \gamma < 0$. As a result,
the spacetime is singular at $u = 0$ for these latter solutions.
However, the singularity is weak for $|\gamma| < 2$, because the distortion exerting on the 
observer is still finite,
\bqn
\lb{3.37}
 \int{d\lambda \int{\hat{R}^{(A)}_{\;\;\;\; (0)(B)(0)} d\lambda}} 
 &\sim & A_{1} \lambda^{2-\gamma} + A_{2} \lambda^{1-\gamma/2} \nb\\
&\sim &  {\mbox{finite}},
\eqn  
as $ \lambda \rightarrow 0 \; (\mbox{or} \; u \rightarrow 0)$ when
$|\gamma| < 2$, where $A_{1}$ and $A_{2}$ 
are finite constants. 

Therefore, for the choice $\epsilon_{a} = +1$ the strong singularities of 
the solutions with $\gamma > 2$ at $u = 0$ in the Einstein frame now 
remain in the string  frame. The singularities of the solutions with $0 < \gamma < 2$ 
are weak in both of the two frames. The solutions with $ -2 < \gamma < 0$ 
is free from singularity in the Einstein frame, while they become singular 
in the string frame, although the nature of the singularities is still weak. 
The solutions with $ \gamma = 0$ and $\gamma \le -2$ are free from singularity 
at $u = 0$ in both of the two frames.  

Note that for $\gamma = 0$ Eq.(\ref{3.33a}) shows that 
\bq
\lb{3.37a}
\hat{R}^{(A)}_{\;\;\;\;(0)(B)(0)} \rightarrow 0,
\eq
as   $u \rightarrow \infty$. Thus, in this case the spacetime singularity 
at $u = \infty$ appearing in the Einstein frame now disappears in the 
$(D+d)$-dimensional string frame, although the null infinity $u = - \infty$ 
still remains singular [cf. Fig. \ref{fig1a}].

When $\epsilon_{a} = - 1$, from Eq.(\ref{3.33}) we find that
\bq
\lb{3.38}
\hat{R}^{(A)}_{\;\;\;\;(0)(B)(0)} = \cases{0, & $\gamma > 2$,\cr
\infty, & $ 2 > \gamma > 0$,\cr
{\mbox{constant}}, & $\gamma = 0$,\cr
\infty, & $- 2 < \gamma < 0$,\cr
{\mbox{finite}}, & $\gamma \le -2$,\cr}
\eq
as $u \rightarrow 0$. It can be shown that in this case the nature of the 
singularities of the solutions remains the same in both of the two frames
for $2 > \gamma \ge  0$ and $\gamma \le -2$, that is, in both frames
it is  
weak for $ 0 < \gamma < 2$, and free of singularities for  $\gamma = 0$
and $\gamma \le -2$.
For $-2 < \gamma < 0$, the solutions are free of singularities  
in the Einstein frame, but singular in the string frame with the nature
of the singularities being still weak.     
For $ 2 < \gamma$, on the other hand, the solutions are free of singularities  
in the string frame, but singular in the Einstein frame with the nature
of the singularities being   strong.   

Similarly, one can show that for $\gamma = 0$ the spacetime singularity 
at $u = -\infty$ appearing in the Einstein frame now disappears in the 
$(D+d)$-dimensional string frame, although the spacetime is still
singular at the null infinity $u = + \infty$ [cf. Fig. \ref{fig1a}].

\subsubsection{$\gamma = 2$} 

When $\gamma = 2$, the corresponding solutions in the Einstein frame
are given by Eqs.(\ref{3.15}) and (\ref{3.16}). The solutions have
a strong singularity at $u = 0$. In the string frame, the corresponding
solutions are given by Eq.(\ref{3.30}) but with
\bqn
\lb{3.39}
\hat{h}(\hat{u}) & =& \frac{a+2\delta}{2(1+a)}\ln\left|\hat{u}\right|,\nb\\
\hat{\Phi}(\hat{u}) & =& \left((1+a)\hat{u}\right)^{\frac{b}{1+a}}, 
\eqn
where
\bq
\lb{3.39a}
\hat{u} \equiv \frac{1}{1+a} u^{1+a} = \cases{0, & $ a > -1$,\cr  
- \infty, & $ a < -1$,\cr}
\eq
as $u \rightarrow 0$. Thus, when $a > -1$ the half plane $u \ge 0$ is mapped
to the half plane $\hat{u} \ge 0$, and the hypersurface $u = 0 \; (u = 
\infty)$ is mapped to the one $\hat{u} = 0 \; (\hat{u} = 
\infty)$. when $a < -1$ the half plane $u \ge 0$ is mapped
to the one $\hat{u} \le 0$, and the hypersurface $u = 0 \; (u = 
\infty)$ corresponds to the one $\hat{u} = -\infty \; (\hat{u} = 
0)$. 

It can be shown that now we have  
\bqn
\lb{3.40}
\hat{R}^{(i)}_{\;\;\;\; (0)(j)(0)} &=&  
- \hat{\gamma}_{0}^{2} \frac{(a+2\delta)(2\delta - a -2)}
{4(1+a)^{2}\hat{u}^{2}}\delta^{i}_{j},\nb\\
\hat{R}^{(a)}_{\;\;\;\; (0)(b)(0)} &=& 
- \hat{\gamma}_{0}^{2} \frac{b(b - a -1)}
{(1+a)^{2}\hat{u}^{2}} \delta^{a}_{b}.
\eqn
Clearly, the spacetime is singular at $\hat{u} = 0$, and the nature
of the singularity is strong, because
\bq
\lb{3.41}
\int{d\lambda\int{\hat{R}^{(A)}_{\;\;\;\;(0)(B)(0)}d\lambda}} \sim \ln{\lambda}
\rightarrow - \infty,
\eq
as $ \lambda \rightarrow 0 \; ({\mbox{or}}\; \hat{u} \rightarrow 0)$. 
Note that the distortion also becomes unbound as $|\hat{u}| \rightarrow    \infty
\; (|\lambda| \rightarrow \infty)$, although the tidal forces vanish there.
 
It should be noted that the above analysis is valid only for $a \not= -1$.
When $a = -1$, we find that
\bq
\lb{3.42}
\omega^{2} = \frac{D+d-2}{4d}, \;\; \; 
b = \frac{D-2}{2d}. 
\eq
The corresponding solutions are given by
\bqn
\lb{3.43}
\hat{h}(\hat{u}) &=& \left(\delta - \frac{1}{2}\right)\hat{u},\nb\\
\hat{\Phi}(\hat{u}) &=& e^{\frac{D-2}{2d} \hat{u}},
\eqn
where $u \equiv  e^{\hat{u}}$.
Then, we find that
\bqn
\lb{3.44}
\hat{R}^{(i)}_{\;\;\;\; (0)(j)(0)} &=&  
- \hat{\gamma}_{0}^{2} \left(\delta - \frac{1}{2}\right)^{2}\delta^{i}_{j}, \nb\\
\hat{R}^{(a)}_{\;\;\;\; (0)(b)(0)} &=& 
- \hat{\gamma}_{0}^{2} \left(\frac{D-2}{2d}\right)^{2}  
\delta^{a}_{b},
\eqn
which are finite (constants). However, at the null infinities 
$\hat{u} = \pm \infty$, which correspond, respectively,  to $u = 0$ and $u = \infty$,
the distortions are still unbound. As a result, the ($D+d$)-dimensional spacetimes  remain
singular on these surfaces. 

Therefore, when $\gamma = 2$   the corresponding Penrose daigram of the ($D+d$)-dimensional 
spacetimes is that of Fig. \ref{fig1a}, where the two hypersurfaces $u = 0$ and $u = \infty$ 
remain singular.

\subsection{ $F'(u) G'(v)  \not= 0$} 

In this case, three classes of solutions were studied in the last section.  To have them 
managible, in this subsection we shall restrict ourselves only to $D = d = 5$. We generalize 
the metric (\ref{3.30}) and using $t=u+v$, $y=u-v$ we arrive at the form:
\bqn
\lb{3.45}
d{\hat{s}}^{2}_{10} &=& \frac{1}{2} e^{2A(t,y)} \left(dt^2-dy^2\right) - e^{2B(t,y)}d\Sigma^{2}_{3} \nb\\
& &  - e^{2C(t,y)}d\Sigma^{2}_{5,z},
\eqn
where $d\Sigma^{2}_{5,z} \equiv {\hat{\gamma}_{ab}\left(z^{c}\right)dz^{a}dz^{b}}\;
(a, b = 1, 2, ..., 5)$, and
\bqn
\lb{3.46}
A &=& \sigma - \frac{5}{3}\beta \phi = - \frac{1}{3} \ln(t) + \kappa^{2}_{5}M -  \frac{5}{3}\beta \phi,\nb\\
B &=& h - \frac{5}{3}\beta \phi = \frac{1}{3} \ln(t) - \frac{5}{3}\beta \phi,\nb\\
C &=&  \beta \phi, \hspace{14pt}
\beta =  \epsilon \sqrt{\frac{3\kappa_{5}^{2}}{40}}, \;\;\;
\epsilon_{a} = \pm 1. 
\eqn

\subsubsection{Class IIa Solutions} 

In this case, substituting the solution (\ref{3.16af}) into Eq.(\ref{3.46}) and setting 
$M_0=\phi_0=0$ without loss of generality, we obtain that
\bqn
\lb{3.47a}
A  &=& \frac{1}{2} \left[\left( c\kappa_5 - \epsilon \sqrt{\frac{5}{24}} \right)^2  
- \frac{7}{8}\right] \ln(t), \nb\\
B  &=& \left(\frac{1}{3} - \epsilon \sqrt{\frac{5}{24}} c\kappa_5\right) \ln(t), \nb\\
C  &=& \epsilon_{a} \sqrt{\frac{3}{40}} c\kappa_5 \ln(t).
\eqn
The corresponding Kretschmann scalar is given by,
\bq
\lb{3.48a}
I_{10} \equiv  R_{ABCD}R^{ABCD} = \frac{\tilde{I}_{10}}{t^{\alpha_0}},
\eq
where
\bqn
\lb{alpha0}
\alpha_0    &=& 2\left(c\kappa_5 - \epsilon_{a} \sqrt{\frac{5}{24}} \right)^2 + \frac{9}{4} > 0, \nb\\
\tilde{I}_{10} &=& \frac{1}{45} \left[9 \chi^{2} \left(40\chi + 143\right)  
	- \epsilon_{a} 52 \sqrt{30} \chi^{3/2}  \left(3 \chi + 4\right)\right.\nb\\
         & & \left. + 80\left(5 \chi +2\right)\right],
\eqn
now with $\chi \equiv c^{2}\kappa^{2}_{5}$.
Clearly, the spacetime is always singular at $t = 0$ for any given $c$, similar to that in 
the 5-dimensional case. Therefore, in the present case, the spacetime singularity remains even
after lifted  from the effective 5-dimensional spacetime to the 10 dimensional bulk.

\subsubsection{Class IIb Solutions} 

In this case, the combination of Eqs.(\ref{3.16ah}) and (\ref{3.46}) yields
\bqn
\lb{3.47b}
A &=& \left[2\left(c\kappa_5 - \epsilon_{a} \sqrt{\frac{5}{96}}\right)^2 
- \frac{7}{16}\right] \ln(t) \nb\\
  & & - \left[\left(c\kappa_5 - \epsilon_{a} \sqrt{\frac{5}{96}}\right)^2 
  - \frac{5}{96}\right] \ln \left (y+\sqrt{y^2-t^2} \right) \nb\\
  & & - \frac{1}{2} \chi^2 \ln \left( y^2 - t^2 \right), \nb\\
B &=&  \left( \frac{1}{3} - \epsilon_{a} \sqrt{\frac{5}{6}} c\kappa_5 \right) \ln(t) \nb\\
  & & - \epsilon_{a} \sqrt{ \frac{5}{24} } c \kappa_5  \ln \left( y+\sqrt{y^2-t^2} \right),  \nb\\
C &=& \epsilon_{a} \sqrt{\frac{3}{10}} c\kappa_5 \ln(t) \nb\\
  & & - \epsilon_{a} \sqrt{\frac{3}{40}} c\kappa_5 \ln\left( y+\sqrt{y^2-t^2}\right),
\eqn
for which we find that
\bq
\lb{3.48b}
I_{10} = \frac{\tilde{I}_{10}}{t^{\alpha_{0}}(y^2-t^2)^{\alpha_{1}}(y+\sqrt{y^2-t^2})^{\alpha_{2}}}, 
\eq
where $\alpha_{0}$ is given by Eq.(\ref{alpha0}),
\bqn
\lb{alpha1}
\alpha_{1} &\equiv& 2 \left(\frac{3}{4} - \chi\right),\nb\\
\alpha_{2} &\equiv& \frac{101}{24} - \left(2 c\kappa_5 - \epsilon_{a} \sqrt{\frac{5}{24}}\right)^2,
\eqn 
and $\tilde{I}_{10} = \tilde{I}_{10}(t,y)$, which is non-zero for $t = 0$ and
$y^{2} = t^{2}$, but its  expression is too complicated to give it here explicitly. 

From the above expression, it can be seen that the spacetime is always singular at $t = 0$, but the
strenght of the singularity for $y > 0$ and $y < 0$ is different, because when $t = 0$ we have
$y+\sqrt{y^2-t^2} = 0$ for $y \le 0$ and $y+\sqrt{y^2-t^2} \not= 0$ for $y > 0$.
In particular, when  $t = 0$ and $y > 0$, we find that
\bq
I_{10} \Big|_{t=0,\; y>0} \simeq \frac{\tilde{I}_{10}}{t^{\alpha_{0}}},
\eq
where
\bqn
\lb{aa}
\tilde{I}_{10} \Big|_{t=0} & = & \frac{256}{45} y^8 \bigg[9\chi^2 \left(160 \chi + 143\right) \nb\\
    & & 
    - \epsilon_{a} 104 \sqrt{30} \chi^{3/2} \left(3 \chi + 1\right) \nb\\
    & & 
    + 10 \left(10 \chi + 1\right)\bigg].
\eqn
On the other hand, when $t = 0$ and $y < 0$, we find that  
\bq
I_{10} \Big|_{t=0, \; y<0} \simeq \frac{y^{\alpha_{2} - 2\alpha_{1}} \tilde{I}_{10}}
{t^{{32}/{3}}}, 
\eq
where $\tilde{I}_{10}$ is still given by Eq.(\ref{aa}).  

Eqs.(\ref{3.48b}) and (\ref{alpha1}) also show that  the spacetime is singular when 
$y^{2} - t^{2} = 0$ for $\chi < 3/4$,   
\bq
I_{10(b)} \Big|_{t^2=y^2} \propto \frac{\tilde{I}_{10}}{(y^2-t^2)^{\alpha_{1}}},
\eq
but now with
\bqn
\lb{4.33}
\tilde{I}_{10} \Big|_{t^2=y^2} &=& \pm\frac{8}{15}t^7\chi \bigg[20(6\chi^{2}-\chi-1) \nb\\
                    & & - 13\epsilon_{a} \sqrt{30} c\kappa_5(2\chi^2-1)\bigg].
\eqn
The corresponding Penrose diagram for $\chi < 3/4$ is given by Fig. \ref{fig2b}.
It is remarkable to note that the solutions with $ 1/2 \le \chi < 3/4$ is not singular in the 
5-dimensional effective theory, as shown explicitly in Sec. III.B.2. 

Since the solution in the 10-dimensional bulk is not singular across the hypersurfaces $y^{2}
= t^{2}$ for $\chi \ge 3/4$, one must extend the solutions beyond these surfaces. The
extension is quite similar to the 5-dimensional case for the ones with $\chi \ge 1/2$.
In particular, for  $3/4 \le \chi < 1$, setting 
\bq
\lb{4.33a}
\bar{u} = (y + t)^{2n},\;\;\;
\bar{v} = (y - t)^{2n},
\eq
where $n$ is given by Eq.(\ref{3.16ak}), one can show that the coordinate singularity at $y^{2} = t^{2}$
disappears in terms of   $\bar{u}$ and $ \bar{v}$. Then, the Penrose diagram for the extended
solutions is given exactly by Fig. \ref{fig2a}. 

When $\chi \ge 1$, the hypersurfaces $y^{2} = t^{2}$ already  represent the null infinities, and
the corresponding Penrose diagram is given by Fig. \ref{fig2b}, but  now the hypersurfaces 
$y^{2} = t^{2}$, represent,
respectively, by the lines $0D$ and $0E$, are non-singular. 
The two regions $I$ and $I'$ are physically disconnected.

\subsubsection{Class IIc Solutions} 

In this case, from Eq.(\ref{3.16am}) we find that
\bqn
\lb{3.47c}
A &=& -\frac{1}{3} \ln(t) - \frac{1}{2} \chi^2 \ln(y^2-t^2)  \nb\\
  & & + \left[\left( c\kappa_5 - \epsilon_{a} \sqrt{\frac{5}{96}} \right)^2  
  - \frac{5}{96}\right] \ln \left( y + \sqrt{y^2-t^2}\right),\nb\\
B &=& \frac{1}{3} \ln(t) - \epsilon_{a} \sqrt{\frac{5}{24}} c \kappa_5 \ln\left(y+\sqrt{y^2-t^2}\right), \nb\\
C &=& \epsilon_{a} \sqrt{\frac{3}{40}} c\kappa_5 \ln \left( y+\sqrt{y^2-t^2} \right),
\eqn
and that 
\bq
\lb{3.48c}
I_{10} = \frac{\tilde{I}_{10}}{t^{{8}/{3}}(y^2-t^2)^{\alpha_{1}}(y+\sqrt{y^2-t^2})^{\alpha_{3}}}, 
\eq
where
\bq
\lb{alpha3}
\alpha_{3} \equiv \frac{91}{24} + \left( 2 c\kappa_5 - \epsilon_{a} \sqrt{\frac{5}{24}} \right)^2 > 0,
\eq
and $\tilde{I}_{10} = \tilde{I}_{10}(t,y)$  is non-zero and finite at $t = 0$, but too complicated
to be written out here. 
From the above expressions we can see that the spacetime is   singular at $t = 0$.
The strength of the singularities once again depends on $y < 0$
and $y > 0$. In partucular, when  $t = 0$ and $y >0$, we find that
\bq
I_{10} \Big|_{t = 0, y > 0} \propto \frac{\tilde{I}_{10}}{t^{8/3}},
\eq
with $\tilde{I}_{10} \simeq   {512}y^7/9$. But for   $t = 0$ and $y < 0$, we find that
\bq
I_{10} \Big|_{t=0, y<0} = \frac{ \tilde{I}_{10}(y-\sqrt{y^2-t^2})^{\alpha_{3}}}
{t^{{8}/{3}+2\alpha_{3}}(y^2-t^2)^{\alpha_{1}}}, 
\eq
where  $\alpha_{3} >0$, as shown by Eq.(\ref{alpha3}). Therefore, we again have a spacetime 
singularity at $t=0$ for $y < 0$, but with more singular strength.

The singular behavior of the spacetime along the hypersurfaces $y = \pm t$ are similar to the last
case. In particular, it is singular  for $\chi < 3/4$,  and corresponding Penrose diagram   is given 
by Fig. \ref{fig2b}. It is interesting to note again that the solutions with $ 1/2 \le \chi < 3/4$ 
is not singular in the 5-dimensional effective theory. 

When $1> \chi \ge 3/4$   the corresponding solutions in the 10-dimensional bulk is not singular 
across the hypersurfaces $y^{2} = t^{2}$, one must extend the solutions beyond these surfaces. The
extension is quite similar to the last case by simply introducing two null coordinates, defined by
Eq.(\ref{4.33a}).  Then, the Penrose diagram for the extended
solutions is given exactly by Fig. \ref{fig2a}. 

When $\chi \ge 1$, the hypersurfaces $y^{2} = t^{2}$ already  represent the null infinities, and
the corresponding Penrose diagram is given by Fig. \ref{fig2b}.

\section{Conclusions and Discussing Remarks}

According to the singularity theorems \cite{HE73}, the 4-dimensional spacetimes are 
generically singular, if  matter filled the spacetimes satisfies certain energy condition(s).
It is a common belief that new physics involved with quantum mechanics will play an important
role    when the spacetime curvature is very high. The new physics gives us hope that
the classical singularities might be finally removed. 

However, it was shown that certain 
singularities can be also removed by simply passing to a higher-dimensional theory of gravity, 
for which spacetime is only effectively four-dimensional below some compactification scale 
 \cite{GHT95,RMS07}. This is because when lifted to higher-dimensions, the conditions required
 by the singularity theorems of Penrose and Hawking do not hold any more. 

In this paper, we have investigated this  problem, by first studying the local and global properties
of the spacetimes in the low dimensional effective theory, and then lifting them to the corresponding
 high dimensional spacetimes. We have shown explicitly that spacetime
singularities may or may not remain after lifted to higher dimensions, depending on the particular
solutions considered. We have also found that there exist cases in which the spacetime singularities of low 
dimensional effective theory occur in different surfaces from those of high dimensional spacetimes. 


\begin{acknowledgements}

One of the authors (AW) would like to express his gratitude to Jianxin Lu for organizing  
the advanced workshop, ``Dark Energy and Fundamental Theory,"  Xidi, Anhui, China,
May 27 - June 5, 2010, supported by the Special Fund for Theoretical Physics from the 
National Natural Science Foundation of China with grant No. 10947203. Part of this work 
was done during the  Workshop. He would also like to thank the other ten participants for 
valuable discussions. The work of AW was supported in part by DOE Grant, DE-FG02-10ER41692.

\end{acknowledgements}

\end{document}